\documentclass[showpacs,floatfix,citeautoscript,twocolumn,prb,superscriptaddress,nobibnotes]{revtex4-1}

\usepackage{graphicx,xcolor}
\usepackage{amsmath}
\usepackage{comment}
\usepackage{textcomp}
\usepackage{inputenc}
\usepackage{xspace}
\usepackage{amssymb}

\newcommand*{\C}{~$^{\circ}$C\xspace}

\begin{document}

\title{High temperature Jahn-Teller distortion and short-range order in CsCuCl$_3$}

\author{Emma A. Pappas}
\affiliation{Department of Physics, University of Illinois Urbana-Champaign, Urbana, IL, 61801, USA}
\affiliation{Materials Research Laboratory, University of Illinois at Urbana-Champaign, Urbana, IL, 61801, USA}

\author{Toby Woods}
\affiliation{School of Chemical Sciences, University of Illinois Urbana-Champaign, Urbana, IL, 61801, USA}

\author{Jue Liu}
\affiliation{Neutron Scattering Division, Oak Ridge National Laboratory, Oak Ridge, Tennessee 37831, United States}

\author{Daniel P. Shoemaker}\email{dpshoema@illinois.edu}
\affiliation{Department of Physics, University of Illinois Urbana-Champaign, Urbana, IL, 61801, USA}
\affiliation{Materials Research Laboratory, University of Illinois at Urbana-Champaign, Urbana, IL, 61801, USA}
\affiliation{Department of Materials Science and Engineering, University of Illinois Urbana-Champaign, Urbana, IL, 61801, USA}

\begin{abstract}
\centerline{\textbf{Abstract}}
CsCuCl$_3$, chiral and Jahn-Teller distorted at room temperature, takes a more symmetric structure above 423~K. Describing this 'simpler' high temperature structure is far from simple and has been the subject of many, sometimes contradicting, studies. Here we reinvestigate the high temperature structure of CsCuCl$_3$, its thermal stability, and its effect on room temperature chirality. \textit{In situ} pair distribution function data from powder neutron diffraction confirms that CsCuCl$_3$ is Jahn-Teller distorted both below and above $T_\mathrm{c}$, and provides a quantitative view of short-range order in the high temperature structure. \textit{In situ} powder x-ray diffraction shows that CsCuCl$_3$ does not undergo any additional structural changes above 423~K and is congruently melting. \textit{In situ} single crystal x-ray diffraction experiments reveal that the phase transition induces domains of mixed handedness in originally homochiral crystals. These findings contribute to a better understanding of phase transitions in Jahn-Teller distorted compounds, and highlight the potential use of phase transitions to control chiral domains.  
\end{abstract}

\maketitle 

\section{Introduction} 

At room temperature, CsCuCl$_3$ has a chiral P6$_1$22 or P6$_5$22 structure with Jahn-Teller distorted CuCl$_6$ octahedra.\cite{schlueter_redetermination_1966} 
The Jahn-Teller distortion arises from the orbital degeneracy of the Cu$^{2+}$ ions,\cite{kroese_high-temperature_1974} resulting in 2 long and 4 short Cu--Cl bonds. The long axis of the CuCl$_6$ octahedra are ordered along the $c$-axis to form a chiral motif. 
Around 423~K, CsCuCl$_3$ undergoes a structural phase transition to a more symmetric, but probably still Jahn-Teller distorted structure, which has been the subject of many studies. 

The structural phase transition of CsCuCl$_3$ was first discovered in 1971 by Natarajan and Prakash,\cite{Natarajan1971_phase_transition} and by Kroese \textit{et al.}\cite{kroese_phase_1971} They investigated the transition using differential thermal analysis, x-ray crystallography, and dielectric constant measurements. Their publications reported different high temperature crystal structures (cubic versus hexagonal), but both indicated that the transition is of first order, which was later confirmed by other studies on the electrical, optical, and thermal expansion properties.\cite{Laiho1973_electrical_optical, hirotsu_optical_1975, sorokin_electrical_2017} 

In 1974, Kroese \textit{et al.}\ reassessed the high temperature structure using single crystal x-ray diffraction.\cite{kroese_high-temperature_1974} They reported a hexagonal polar structure with space group P6$_3$mc. In their model, the Cu$^{2+}$ ion is displaced along the $c$-axis, away from the center of the surrounding chlorine octahedron, resulting in a ferroelectric structure. The chlorine ions also have some anisotropic thermal motion along the $c$-axis. Kroese and Maaskant continued their study of the phase transition by assuming an idealized high temperature structure in the space group P6$_3$/mmc. They found a softening vibration mode at $k=(0,0,2\pi/3c)$.\cite{kroese_relation_1974} 

In 1975 and 1977, Hirotsu\cite{hirotsu_optical_1975, hirotsu_jahn-teller_1977} also showed that the $k=\pm(0,0,2\pi/3c)$ phonon mode relates the high and low temperature phases and investigated the optical properties of CsCuCl$_3$ around the phase transition. They found that the optical rotation is not recovered upon heating and cooling and proposed that the transition induces enantiomorphic domains smaller than those originally present in the pristine crystal.\cite{hirotsu_optical_1975}

In 1981, many publications introduced new information about the high temperature structure, contradicting or nuancing the results of previous studies. The far-infrared and Raman spectroscopy experiments of Petzelt \textit{et al.}\ showed that, above the phase transition, CsCuCl$_3$ is non-polar with a high degree of disorder.\cite{petzelt_farinfrared_1981} They found that the phase transition has a soft mode with an infrared-active phason component. Crama redetermined the high temperature structure from  single crystal x-ray diffraction experiments.\cite{crama_jahnteller_1981} Multiple models taking into account the presence of a local Jahn-Teller effect were considered. The model with 3 split chlorine sites was determined to be the most probable and no softening of the $k=\pm(0,0,2\pi/3c)$ mode was found. Electron paramagnetic resonance (EPR) studies by Tanaka \textit{et al.}\ suggested that static local distortions persist in the high temperature phase along with intra-chain interactions.\cite{Tanaka_1981_EPR} Intra- and inter-chain interactions are also discussed in other works,\cite{tazuke_magnetic_1981, tanaka_structural_1986, tanaka_structural_1986_part2} where the theory supports the idea that the CuCl$_6$ octahedra are Jahn-Teller distorted and orientationally disordered at high temperature.\cite{tanaka_structural_1986} 

In 1985, Tanaka \textit{et al.}\ conducted additional EPR experiments.\cite{tanaka_electron_1985} They once again proposed that the CuCl$_6$ octahedra are Jahn-Teller distorted at high temperatures and showed that the distortions have long relaxation times. In 1986, Haije and Maaskant did magnetic susceptibility measurements on centimeter size crystals.\cite{haije_magnetic_1986} They observed a transition from a static Jahn-Teller effect to a cooperative dynamic Jahn-Teller effect occurring in domains. That same year, Graf \textit{et al.}\ performed neutron scattering experiments and claimed that the diffuse scattering at high temperature is incompatible with intra-chain correlations and could be interpreted as dense impurity Huang scattering (diffuse scattering coming from the elastic strain fields of defects, or in this case, Jahn-Teller distortions).\cite{graf_jahn-teller_1986} In 1987, Schotte showed that the neutron diffraction diffuse scattering was not due to short-range correlations.\cite{schotte_theory_1987} The dense Huang scattering theory was continued in 1988 by Schotte \textit{et al.}\cite{schotte_elastic_1989} Additional neutron scattering experiments and theoretical work were done by Graf \textit{et al.}\ in 1989.\cite{graf_quasi-elastic_1989} They suggested that the first-order order-disorder transition could be triggered by an evolving second-order displacive transition far below the observed transition temperature. In 1997, another neutron experiment by Förster \textit{et al.}\ saw some phonon softening at small $q$-values.\cite{forster_phonon_1997}

Many other works have investigated the properties and phase transition of CsCuCl$_3$. For example, Maaskant and Haijie (1986) and Maaskant (1995) discussed the phenomenon of structural resonance,\cite{maaskant_jahn-teller-induced_1986, maaskant_helices_1995} and Fernández \textit{et al.}\ (1976), Vasudevan \textit{et al.}\ (1979), and Bazán \textit{et al.}\ (2009) reported additional high temperature phase transitions seen in measurements such as differential thermal analysis and differential scanning calorimetry.\cite{fernandez_new_1976, VASUDEVAN197944, bazan_reduction_2011}

It is easy to see that the research on CsCuCl$_3$ contains many contradictions. 
The non-polarity of the high temperature structure, the presence of soft modes and intra-chain correlations, and the nature of the first-order transition have all been questioned. Our work aims to clarify certain assertions about CsCuCl$_3$. In this paper, we reinvestigate the high temperature structure of CsCuCl$_3$, its thermal stability, and the effect of the phase transition on chirality. 
First, we present new temperature dependent pair distribution function data from powder neutron diffraction. Our results confirm that CsCuCl$_3$ is Jahn-Teller distorted above the phase transition and quantify the Cu--Cl bond lengths.
We show that these distortions are correlated over multiple polyhedra. 
Second, we use \textit{in situ} powder x-ray diffraction to search for additional structural phase transitions. No additional structural phase transitions are found up to the melting point of the compound. Third, we use \textit{in situ} and \textit{ex situ} single crystal x-ray diffraction to show that the phase transition induces chiral domains.

\section{Methods}

\subsection{Synthesis of CsCuCl$_3$}

Single crystals of CsCuCl$_3$ were obtained by aqueous precipitation from solutions containing CuCl$_2$ and CsCl. Various synthesis conditions have been reported in the literature and it is often mentioned that an excess of CuCl$_2$ is needed to avoid the precipitation of Cs$_2$CuCl$_4$.\cite{oconnor_preparation_1970, adachi_helical_1980, koiso_determination_1996, soboleva_growth_2008, kousaka_crystal_2014} We used crystals from three different batches of precipitation for our measurements. Each solution was prepared using granular powders of CuCl$_2$ (99\% Acros) and CsCl (99.9\% Sigma Aldrich) dissolved in deionized water with a 2:1 molar concentration. The solutions were filtered and left to evaporate for a few days to a few weeks in partially covered petri dishes or beakers. Evaporation occurred at room temperature on a benchtop or in a fumehood. Crystals were rinsed with ethanol or deionized water before being used for measurements. Additional synthesis details for the crystals used in our measurements are available in the Supporting Information.




We performed other precipitations with various precursor concentrations and temperatures, and with and without filtering, seed crystals and surfactants. All synthesis conditions resulted in the precipitation of CsCuCl$_3$, as confirmed by powder x-ray diffraction, although some reactions also precipitated Cs$_3$Cu$_3$Cl$_8$OH.\cite{van_well_mixed_2020} Cs$_3$Cu$_3$Cl$_8$OH crystals are recognizable by their rectangular shape and dark orange-red color.
None of our reactions precipitated Cs$_2$CuCl$_4$. The morphology of our black CsCuCl$_3$ crystals is reminiscent of a rugby ball and reflection x-ray diffraction indicated that the long axis is the $c$-axis, in agreement with Soboleva \textit{et al.}\ and Cui \textit{et al.}\cite{soboleva_investigation_1976, cui_synthesis_2020} We sometimes noticed liquid inclusions in certain crystals, which was also reported by Tanaka \textit{et al.}\cite{tanaka_electron_1985} Additional observations about the results of our various precipitation experiments can be found in the Supporting Information.

\subsection{Sample Characterization and Measurements}

Portions of the samples were crushed and analyzed using powder x-ray diffraction (PXRD) on a Bruker D8 diffractometer with a Mo x-ray source in transmission geometry. The sample for the \textit{in situ} PXRD measurements was diluted with ground quartz before being sealed in an evacuated quartz capillary. The heating rate was 60~K/min. The fixed temperature scans lasted 93 minutes and were preceded by a 7-minute hold. The crystal structures were refined using the Rietveld method with GSAS-II\cite{GSAS2}. Representative fits to the PXRD data are available in the Supporting Information. 

The chemical composition of some of the samples was analyzed using a scanning electron microscope (SEM) with energy dispersive x-ray spectroscopy (EDS). Representative results can be found in the Supporting Information along with details about the measurements. 

Temperature dependent neutron diffraction and pair distribution function (PDF) data were collected at the NOMAD instrument at Oak Ridge National Laboratory.\cite{NOMAD,ORNL} Two samples of CsCuCl$_3$ were measured and gave very similar data, so only one dataset is presented in this paper. Approximately 250~mg of powdered CsCuCl$_3$ was loaded into a 3~mm diameter thin-walled quartz capillary. An Oxford Instruments argon cryostream was used for temperature control. The sample was first cooled to 100~K and then heated to the target temperature at a rate of 6~K/min, where it was equilibrated for 5~min prior to data collection. At each temperature point, two 24~min scans were collected and summed to improve counting statistics. The diffraction data were background-subtracted using scattering data collected from an empty quartz capillary at the corresponding temperatures. The resulting scattering signals were then normalized to the scattering from a 6~mm diameter vanadium rod to correct for detector efficiency. A second-order polynomial function was used to empirically model the Placzek correction and obtain the arbitrarily normalized total scattering function, $S(Q)$. A sine Fourier transform was then performed using a $Q_\textrm{max}$ cutoff of 25~\AA$^{-1}$. Fourier filtering was applied to remove contributions from uncorrected multiple and inelastic scattering. A near-absolute $g(r)$ was obtained by fine-tuning the atomic number density such that $g(r)$ approached zero for interatomic distances smaller than 2~\AA. The reduced data $G(r) = 4\pi r\rho_0(g(r)-1)$ was fitted using PDFgui.\cite{Farrow_2007} Values for $Q_\textrm{damp}$ and $Q_\textrm{broad}$ were obtained from fitting calibration data collected on silicon at 300~K. The calibration data was Fourier transformed using a $Q_\textrm{max}$ cutoff of 40~\AA$^{-1}$ and $G(r)$ was fitted from 0.5 to 50~\AA. The CsCuCl$_3$ PDF data was fitted from 0.5 to 20~\AA. In each fit, the data scale factor and delta1 (for data at 400, 450 and 500~K) or delta2 (for data at 100 and 300~K) were refined. The refinement conditions for the lattice parameters, atom positions, and thermal displacement parameters are specified in the discussion. 

Differential scanning calorimetry (DSC) measurements were made on a Waters TA Instruments Discovery 2500 Differential Scanning Calorimeter. A few small crystals (approximately 50~mg) were placed in  pressed aluminum pans and heated to 573~K at a rate of 10~K/min. The measurement was performed on heating, cooling, and heating again, and was repeated on a second sample of similar mass. 

Fragments of CsCuCl$_3$ crystals were used for single crystal x-ray diffraction (SCXRD). Intensity data were collected on a Bruker D8 Venture kappa diffractometer equipped with a Photon II (for the room temperature experiments) or Photon III (for the \textit{in situ} experiments) CPAD detector. An I$\mu$s microfocus source provided the Mo K$\alpha$ radiation ($\lambda$ = 0.71073~\AA) that was monochromated with multilayer mirrors. The sample was mounted on a 0.3~mm nylon loop with the minimal amount of Paratone-N oil. Data was collected as a series of $\phi$ and/or $\omega$ scans. For the \textit{in situ} experiments, temperature control was achieved using a stream of N$_2$(g) provided by an Oxford Cryosystems Cryostream 1000. Initial data collection was performed at 298 K. The crystal was subsequently heated at 150~K/hr to 450~K. A 15~minute hold time at 450~K was used prior to commencement of data collection at 450~K. Upon completion of data collection at 450~K, the crystal was cooled at 150~K/hr back to 298~K. Following a 15~minute hold, a post-heating data set was collected at 298~K. The collection, cell refinement, and integration of intensity data was carried out with the APEX3 (room temperature) or APEX5 (\textit{in situ}) software.\cite{apex3} The absorption corrections were performed by multi-scan methods using SADABS.\cite{krause_comparison_2015} The structures were solved with dual space methods using SHELXT version 2018/2.\cite{sheldrick_shelxt_2015A} The structures were refined with the full-matrix least-squares routine of SHELXL version 2019/3.\cite{sheldrick_crystal_2015C} Data collection at a given temperature took approximately 15 minutes.

The crystal structure drawings were produced with VESTA.\cite{Momma:db5098}

\begin{figure*}[!htbp]
    \centering
    \begin{minipage}[t]{\textwidth}
        \begin{minipage}[t]{0.3\textwidth}
            \centering
            \Large Chiral P6$_1$22 
        \end{minipage}
        \hfill
        \begin{minipage}[t]{0.3\textwidth}
            \centering
            \Large HT P6$_3$mc 
        \end{minipage}
        \hfill
        \begin{minipage}[t]{0.3\textwidth}
            \centering
            \Large  HT P6$_3$/mmc 
        \end{minipage}
    \end{minipage}
    
    \centering
    \hspace{-0.8cm}
    \begin{minipage}[b]{0.3\textwidth}
        \centering
        \begin{minipage}[b]{0.4\linewidth}
            \centering
            (a)
            \includegraphics[height=6cm]{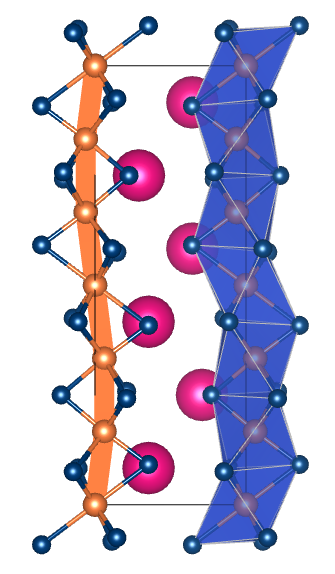}
        \end{minipage}
        \hfill
        \begin{minipage}[b]{0.4\linewidth}
            \centering
            (b)
            \begin{minipage}[t]{\linewidth}
                \centering
                \includegraphics[height=2.3cm]{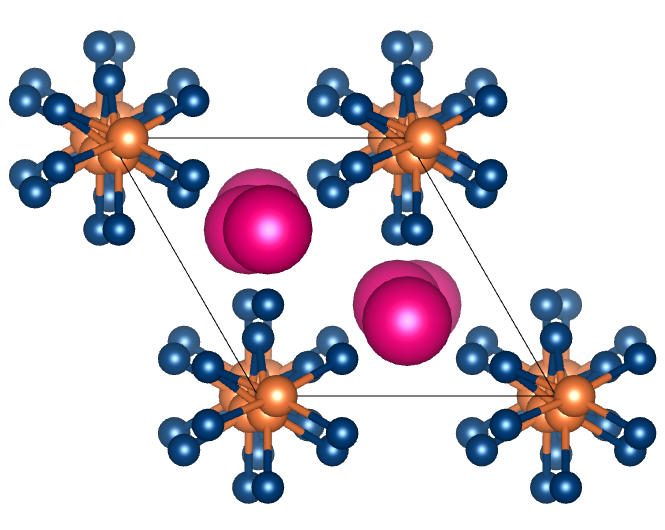}
            \end{minipage}
        
            \vspace{0.3cm}
            (c)
            \begin{minipage}[t]{\linewidth}
                \centering
                
                \includegraphics[height=2.5cm]{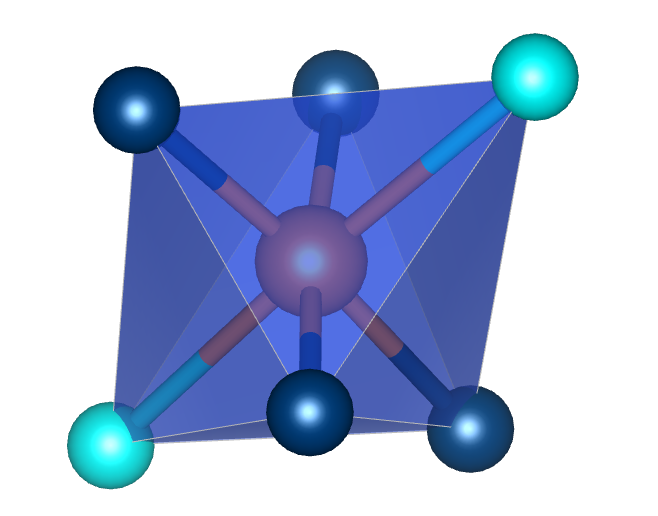}
            \end{minipage}
        \end{minipage}
    
    \end{minipage}
    \hspace{0.7cm}
    \begin{minipage}[b]{0.3\textwidth}
        \centering
        
        \begin{minipage}[b]{0.4\linewidth}
            \centering
            (d)
            \includegraphics[height=6cm]{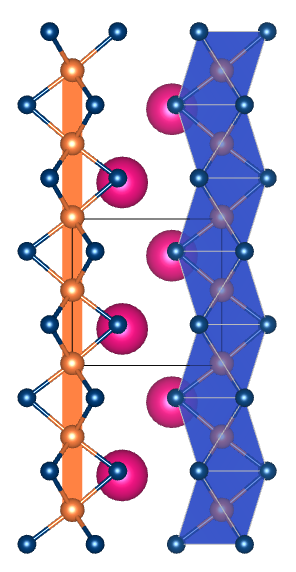}
        \end{minipage}
        \hspace{0.55cm}
        \begin{minipage}[b]{0.4\linewidth}
            \centering
            (e)
            \begin{minipage}[t]{\linewidth}
                \centering
                \includegraphics[width=3cm]{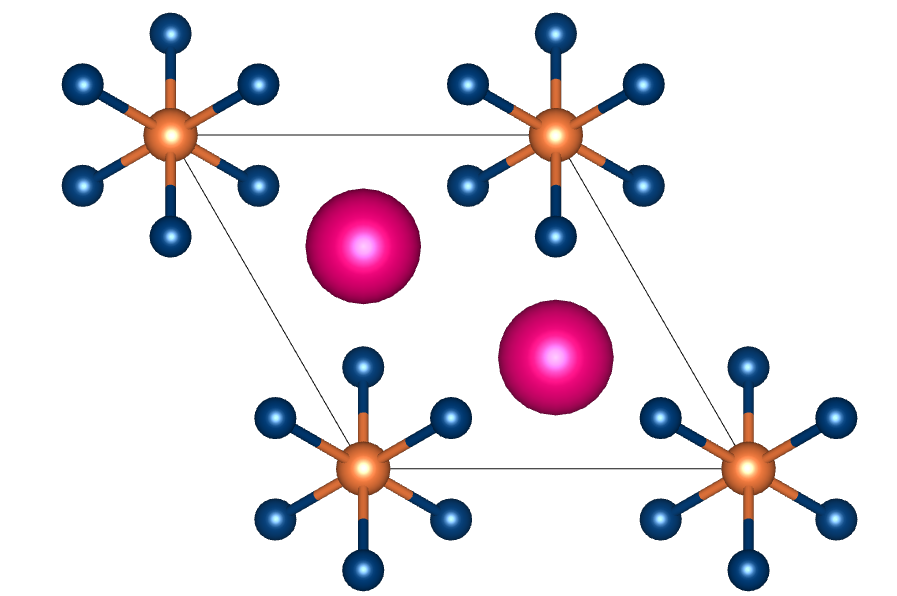}
            \end{minipage}
        
            \vspace{0.3cm}
            (f)
            \begin{minipage}[t]{\linewidth}
                \centering
                \includegraphics[height=2.5cm]{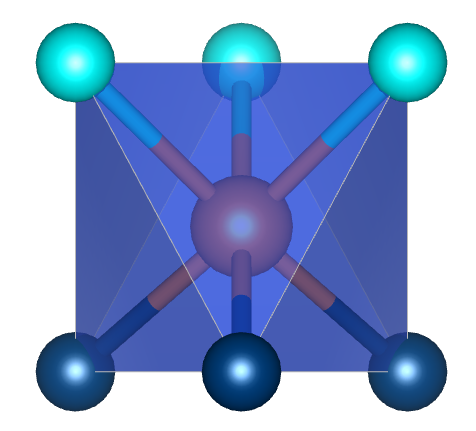}
            \end{minipage}
        \end{minipage}
    
    \end{minipage}
    \hspace{0.7cm}
    \begin{minipage}[b]{0.3\textwidth}
        \centering
        
        \begin{minipage}[b]{0.4\linewidth}
            \centering
            (g)
            \includegraphics[height=6cm]{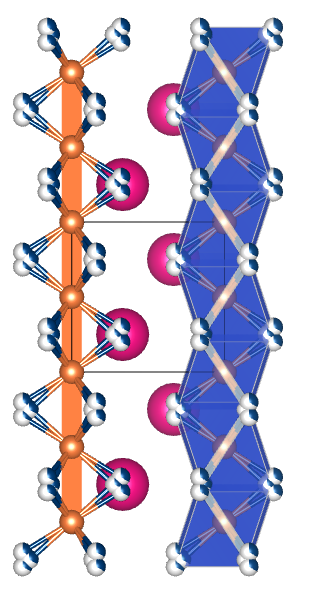}
        \end{minipage}
        \hspace{0.55cm}
        \begin{minipage}[b]{0.4\linewidth}
            \centering
            (h)
            \begin{minipage}[t]{\linewidth}
                \centering
                \includegraphics[width=2.9cm]{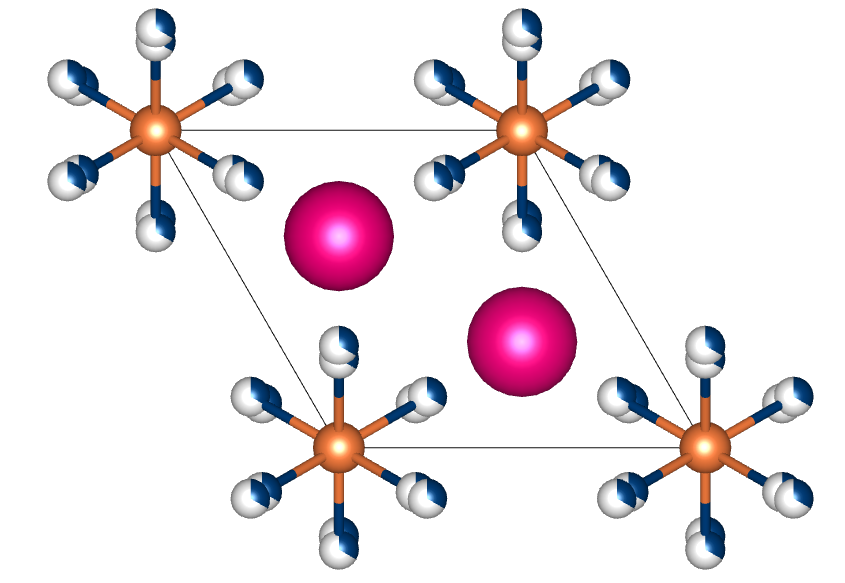}
            \end{minipage}
        
            \vspace{0.3cm}
            (i)
            \begin{minipage}[t]{\linewidth}
                \centering
                \includegraphics[height=2.5cm]{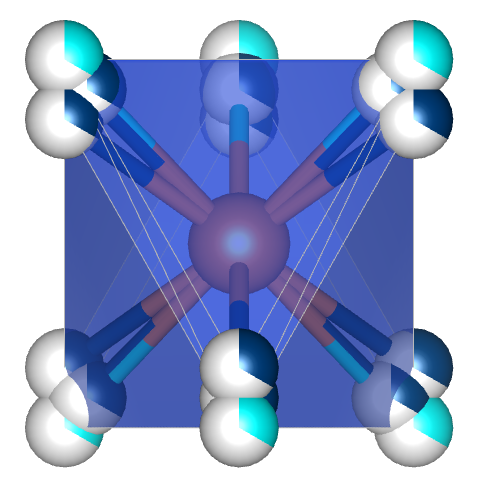}
            \end{minipage}
        \end{minipage}
    
    \end{minipage}
    \caption{Crystal structures of CsCuCl$_3$ with the Cs, Cu and Cl atoms in pink, orange and blue, respectively. Black lines outline single unit cells. (a)-(c) Room temperature P6$_1$22 structure, (d)-(f) High temperature single-site P6$_3$mc structure, (g)-(i) High temperature split-site P6$_3$/mmc structure. Views of the structures along the $a$-axis are shown in (a), (d) and (g). The CuCl$_6$ octahedra are displayed on the right half of the structures. On the left half, orange highlights emphasize the Cu spiral found in the chiral P6$_1$22 structure, but absent from the achiral P6$_3$mc and P6$_3$/mmc structures. Views of the structures along the $c$-axis are shown in (b), (e) and (h). In (c), (f) and (i), individual CuCl$_6$ octahedra are shown. Cu--Cl bonds 2.5~\AA~and longer are marked by light blue copper ions. }
    \label{fig:Structures}
\end{figure*}

\section{Results and Discussion}

\subsection{Pair Distribution Function Analysis: Local Structure of CsCuCl$_3$}

Below 423~K (or 150~\C), CsCuCl$_3$ crystallizes in the P6$_1$22 or P6$_5$22 enantiomorphous structure pair. The chiral nature of the structures is noticeable in the positions of the cesium and copper ions, which form helices along the $c$-axis (see Fig.~\ref{fig:Structures}(a)-(b)). The helix formed by the copper ions in the P6$_1$22 structure is highlighted in orange in Fig.~\ref{fig:Structures}(a). The chlorine ions form Jahn-Teller distorted octahedra around the copper ions, giving rise to  4 short and 2 long Cu--Cl bonds of lengths 2.281(4), 2.354(4), and 2.776(4)~\AA ~(two bonds of each length per copper ion). The CuCl$_6$ octahedra are said to have pseudo D$_{4h}$ symmetry.\cite{crama_jahnteller_1981} Upon warming above 423~K, CsCuCl$_3$ undergoes a structural change to a lower symmetry structure. The low temperature structure has a unit cell 3 times longer along $c$ than the high temperature structure. 

Discrepancies in the literature indicate that previous studies have been unable to accurately describe the chlorine positions in the high temperature phase. Two models have been proposed: a single-site P6$_3$mc structure\cite{kroese_high-temperature_1974} and a split-site P6$_3$/mmc structure\cite{crama_jahnteller_1981}. Both models are achiral as can be seen by the absence of the Cs and Cu spirals found in the low temperature chiral structure (see Fig.~\ref{fig:Structures}(a), (d) and (g)). The P6$_3$mc structure, shown in Fig.~\ref{fig:Structures}(d)-(f), reports large thermal ellipsoids to account for the uncertainty in the Cl positions. In this average structure, each copper ion makes 3 Cu--Cl bonds of 2.39~\AA ~and 3 Cu--Cl bonds of 2.51~\AA, resulting in a C$_{3v}$ site symmetry. The P6$_3$/mmc structure, seen in Fig.~\ref{fig:Structures}(g)-(i), places the Cl ions at 3 possible positions with equal probability to try to account for the presence of a local Jahn-Teller distortion. These split-sites give rise to possible Cu--Cl bond distances of 2.33, 2.36 or 2.72~\AA, and an average D$_{3d}$ site symmetry for the copper ion. 

\begin{figure}
    \centering
    \includegraphics[width=\columnwidth]{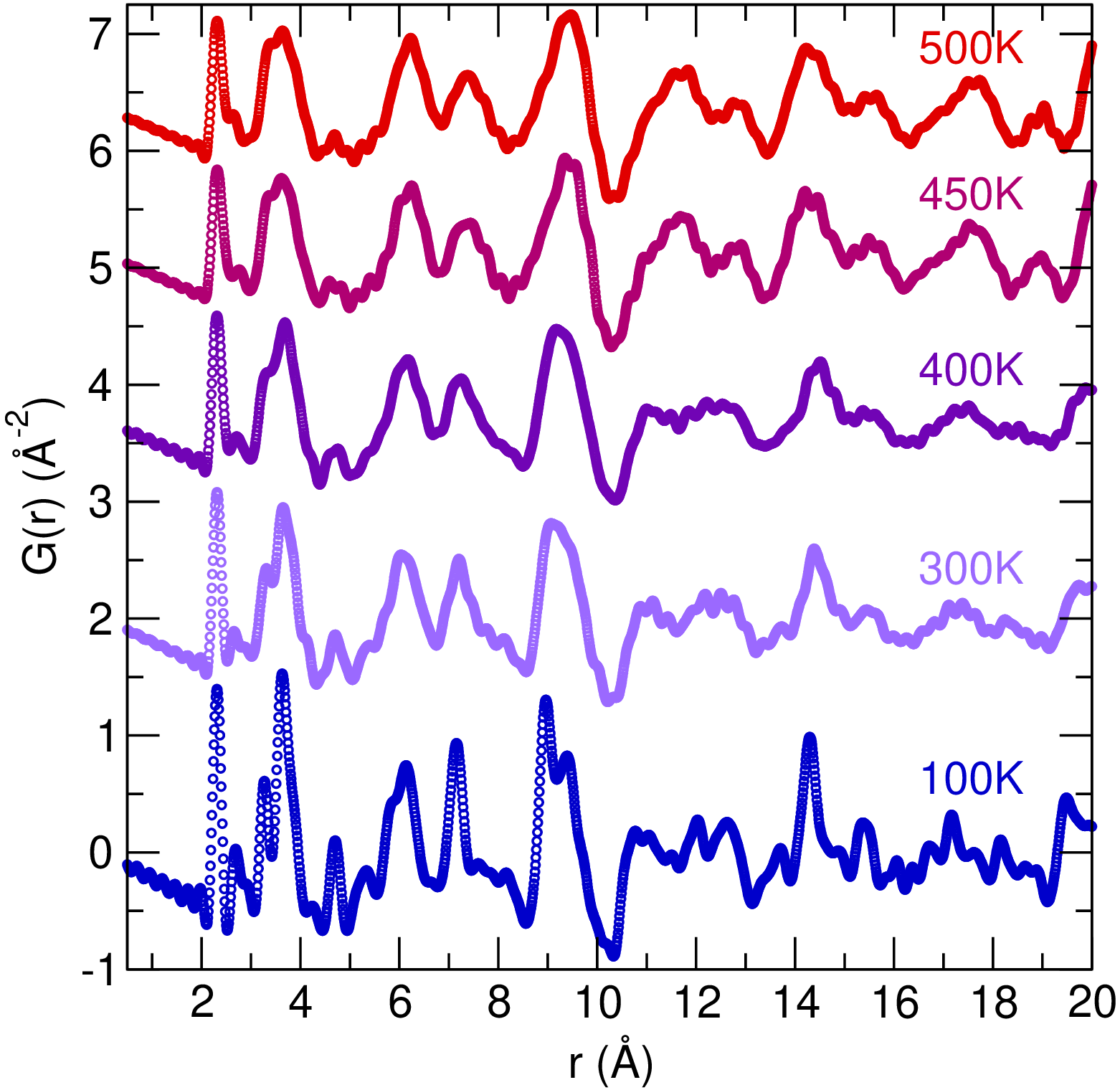}
    \caption{Pair distribution function data collected on CsCuCl$_3$ at various temperatures. Below 423~K, the sample is in the low temperature chiral structure, while above 423~K, the sample takes an achiral structure. The tall first peak and its smaller shoulder peak between 2 and 3~\AA~correspond to the Jahn-Teller distorted Cu--Cl bonds and are visible at all temperatures.} 
    \label{fig:PDFdata}
\end{figure}

\begin{figure}
    \centering
    \includegraphics[width=\columnwidth]{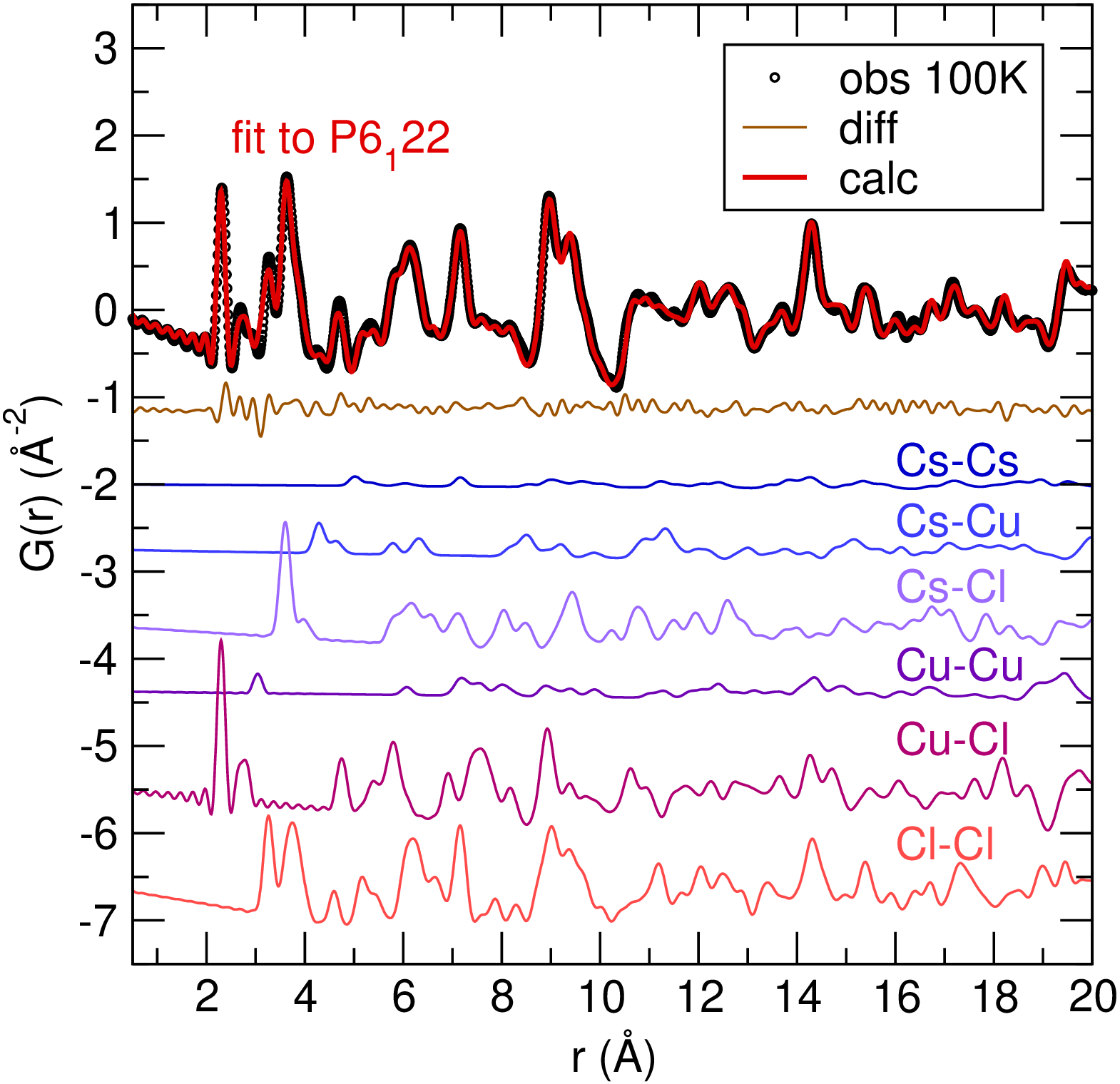}
    \caption{The P6$_1$22 structure of CsCuCl$_3$ is fit (red curve) to the PDF data at 100~K (black circles). The residuals (brown) and the contributions of each bond to the fit (labeled) are displayed below the fitted data. This fit verifies the chiral P6$_1$22 structure of CsCuCl$_3$  below the structural transition at 423~K.}
    \label{fig:PDF_fit_100K}
\end{figure}

\begin{figure}
    \centering
    \includegraphics[width=\columnwidth]{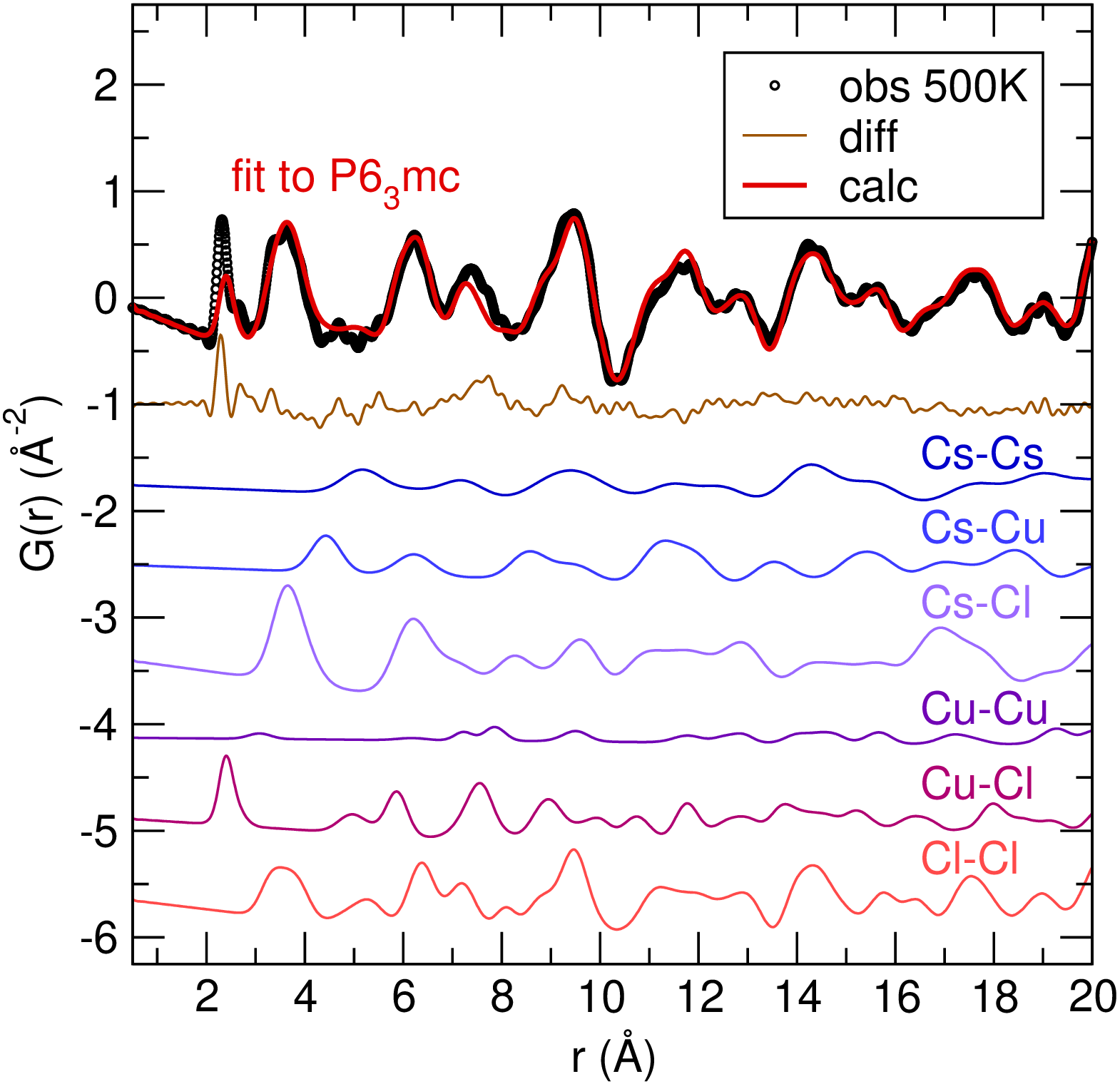}
    \caption{The single-site P6$_3$mc structure of CsCuCl$_3$ is fitted (red curve) to the PDF data at 500~K (black circles). The residuals (brown) and the contributions of each bond to the fit (labeled) are displayed below the fitted data. Significant misfits of the Cu--Cl and Cl--Cl bonds can be observed below 4~\AA\  in the difference curve.}
    \label{fig:PDF_fit_disordered}
\end{figure}

\begin{figure}
    \centering
    \includegraphics[width=\columnwidth]{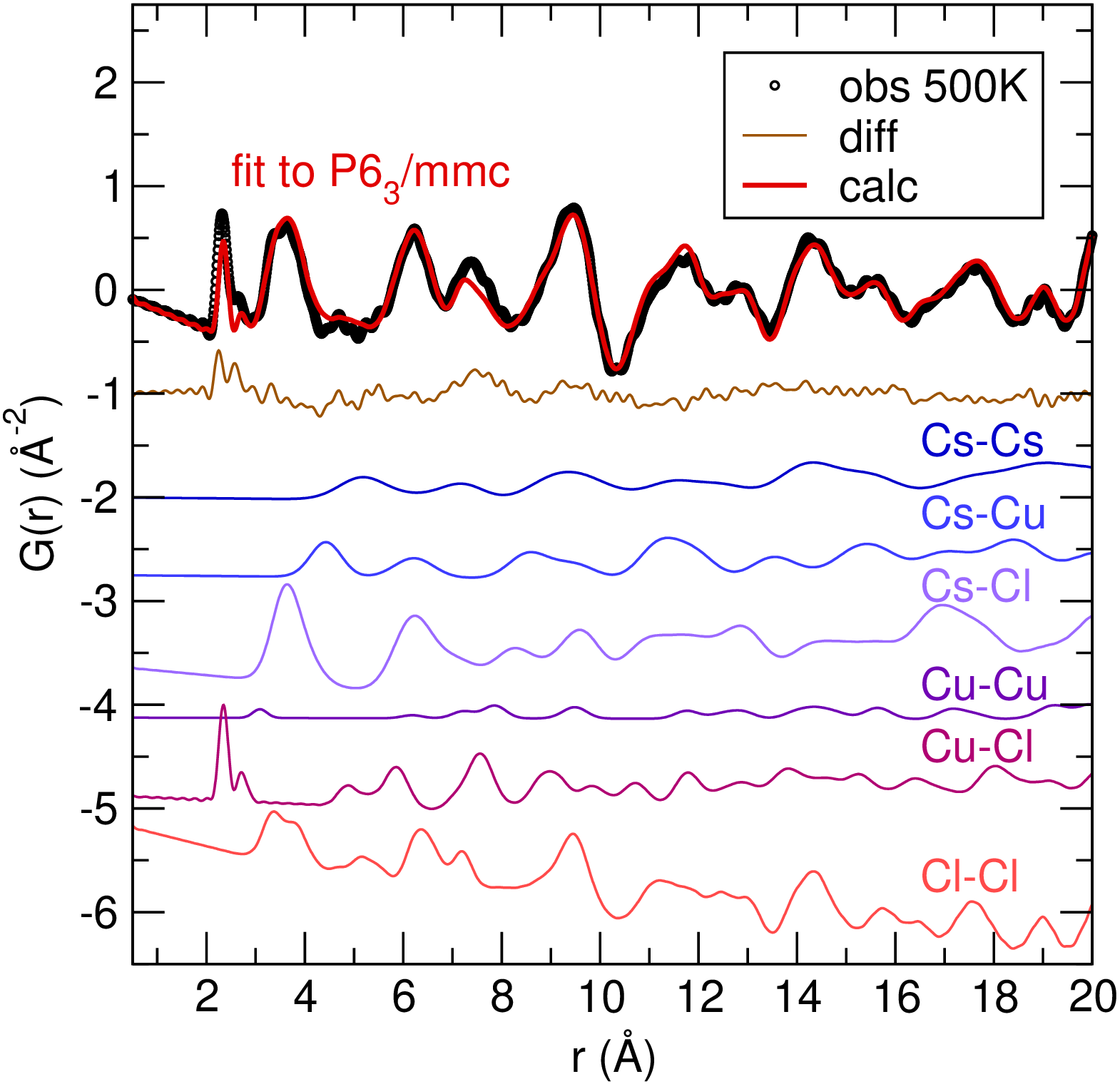}
    \caption{The split-site P6$_3$/mmc structure of CsCuCl$_3$ is fitted (red curve) to the PDF data at 500~K (black circles). The residuals (brown) and the contributions of each bond to the fit (labeled) are displayed below the fitted data. As in Fig.\ref{fig:PDF_fit_disordered}, the fit is poor at low $r$. }
    \label{fig:PDF_fit_split-site}
\end{figure}

To investigate the local structure of the high temperature phase, we collected pair distribution function (PDF) data at 100, 300, 400, 450 and 500~K (see Fig.~\ref{fig:PDFdata} and the larger Fig.~S5 in the Supporting Information). The data at 100, 300 and 400~K, below the 423~K structural transition, display similar features, which broaden as the temperature increases due to thermal vibrations. At 450 and 500~K, the peaks around 2.5, 7.5, 11.5 and 17.5~\AA~have different intensity profiles compared to the low temperature data. These differences highlight the fact that the material has undergone a structural change. The structural transition is also noticeable in powder x-ray diffraction data (see Section~\ref{sec:stability} and the fits at 25, 200 and 400\C in the Supporting Information). 
A feature of interest in the PDF data (Fig.~\ref{fig:PDFdata}) is the tall first peak and its smaller shoulder peak between 2 and 3~\AA, present at all temperatures. For every CsCuCl$_3$ structure, this length scale corresponds to the Cu--Cl bonds (see the bond contributions to the fit in Fig.~\ref{fig:PDF_fit_100K}, \ref{fig:PDF_fit_disordered} and \ref{fig:PDF_fit_split-site}). At low temperature, this feature is explained by the 4 short and 2 long Cu--Cl bonds created by the Jahn-Teller distortion. The fact that this feature persists in the 450 and 500~K datasets is a first indication that the CuCl$_6$ octahedra are Jahn-Teller distorted even at high temperatures. 

As expected, the P6$_1$22 structure is a good fit to the pair distribution function data collected at 100, 300 and 400~K, which is below the structural transition at 423~K. Figure \ref{fig:PDF_fit_100K} shows the fit at 100~K. The refinements include the lattice parameters and atomic displacement parameters constrained by symmetry, with the atom positions fixed to those obtained from single crystal diffraction data.\cite{schlueter_redetermination_1966} All fits of the P6$_1$22 structure to the low temperature data have small residuals and weighted R-values between 0.11 and 0.15. 

Fits of the single-site P6$_3$mc and split-site P6$_3$/mmc models to the 500~K data are shown in Fig.~\ref{fig:PDF_fit_disordered} and \ref{fig:PDF_fit_split-site}. Fits of the 450~K data give similar results. For both models, the lattice parameters and atomic displacement parameters are constrained and refined by symmetry, while the atom positions known from single crystal diffraction data\cite{kroese_high-temperature_1974, crama_jahnteller_1981} are not refined. To fit the P6$_3$/mmc split-site structure, we use a 2x2x2 supercell with random chlorine site selection. Both models result in mis-fits of the first peak, its shoulder and the peak around 7.5~\AA. Bond contributions indicate that these problems are mostly associated with the Cu--Cl and Cl--Cl bond distances. 

As seen in Fig.\ref{fig:PDF_fit_disordered}, the P6$_3$mc model is a particularly poor match to the data between 2 and 3~\AA. Its unique Cl site gives rise to too-similar Cu--Cl bond lengths of 2.39 and 2.51~\AA, creating a single broad peak instead of two. The poor fit of the P6$_3$mc structure is to be expected, since it does not model the Jahn-Teller distortion visible in the data. The C$_{3v}$ symmetry of the Cu site gives 3 medium-short and 3 medium-long Cu--Cl bonds, instead of the 4 short and 2 long Cu--Cl bonds expected from the Jahn-Teller distortion. 

The split Cl sites of the P6$_3$/mmc model result in a slightly better fit, as the possible Cu--Cl distances of 2.33, 2.36 or 2.72~\AA~create the required two peaks between 2 and 3~\AA~(see Fig.\ref{fig:PDF_fit_split-site}). However, the random chlorine site selection does not necessarily give 2 long and 4 short Cu--Cl bonds arranged in the pseudo D$_{4h}$ symmetry expected from the Jahn-Teller distortion. This symmetry mismatch could contribute to the poor fit of the P6$_3$/mmc structure at low $r$.

Overall, both high temperature models display significant misfits at short ranges and only adequately fit the data at long ranges. This shows that both models are representations of the average high temperature structure and incorrectly describe the precise positions of the chlorine ions. It is rather surprising that the P6$_3$/mmc structure is such a poor fit at short ranges considering that the split-sites attempt to take into account the presence of a Jahn-Teller distortion at high temperature. The poor fit could be explained by short-range order, which we examine subsequently.

\subsection{Pair Distribution Function Analysis: Short-Range Ordering}

To confirm that Jahn-Teller distorted CuCl$_6$ octahedra accurately describe the local high temperature structure and to verify the existence of a short-range order, we fit the CsCuCl$_3$ room temperature structure over different ranges of $r$. Fits are performed in the range $r=0.5$~\AA~to $r_\textrm{max}$, where $r_\textrm{max}$ varies from 5 to 20~\AA~by steps of 1~\AA. The atom positions of the room temperature structure are refined and constrained by symmetry. To reduce the number of free parameters, we consider one isotropic thermal displacement parameter per element. We also fix the lattice parameters of all $r_\textrm{max}$ fits to the values obtained for $r_\textrm{max}=20$~\AA. The fits were sometimes sensitive to the initial guesses for the refined parameters, so for consistency, the initial guesses for a fit at $r_\textrm{max}=r$ are the results of the fit at $r_\textrm{max}=r+1$. Fits at different $r_\textrm{max}$ values are performed on all datasets for comparison purposes. 

Fits with low $r_\textrm{max}$ values are only influenced by short-range structural information. As $r_\textrm{max}$ increases, information about the long-range structure increases and the fits have to accommodate both the local and average structures, which differ at high temperatures due to the lack of long-range order. The resulting atom positions should then gradually tend to those of the average structure, given by the P6$_3$mc or P6$_3$/mmc model. A similar procedure has been employed to demonstrate the presence of local order in LaMnO$_3$\cite{qiu_orbital_2005}, which has stronger correlations and larger locally ordered domains than CsCuCl$_3$, as we discuss below.

\begin{figure}[t]
    \centering
    \includegraphics[width=\columnwidth]{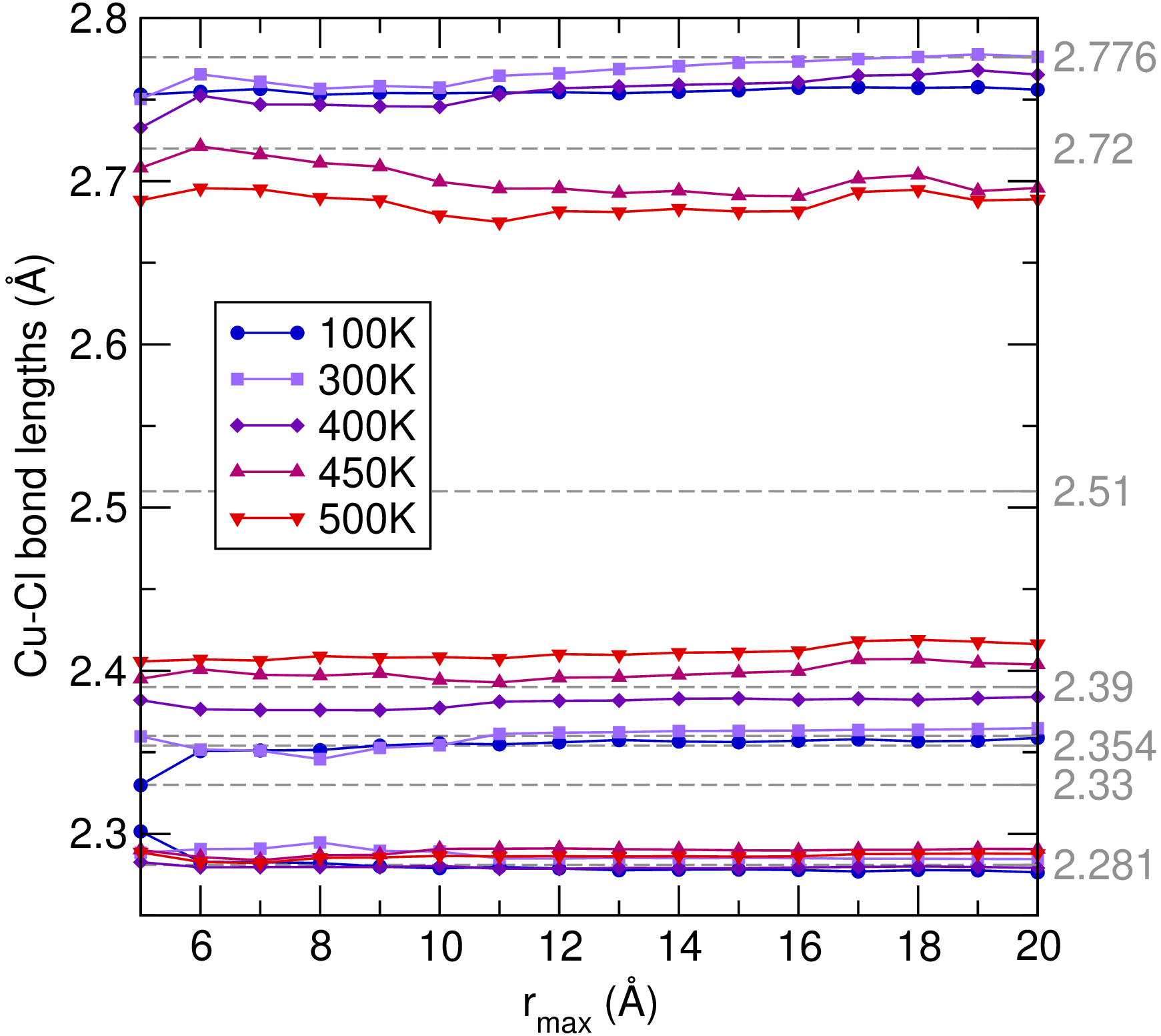}
    \caption{Length of the three unique Cu--Cl bonds as the P6$_1$22 structure is fitted to PDF datasets with different $r_{max}$ cutoff. Fitted values for datasets at 100, 300, 400, 450 and 500~K are shown with blue circles, lilac squares, purple diamonds, burgundy up triangles and red down triangles, respectively. Grey dotted lines indicate the reported values for the P6$_1$22, P6$_3$mc and P6$_3$/mmc structures (see text for more details). All temperatures and all $r_\textrm{max}$ display one long and two short Jahn-Teller distorted bonds.}
    \label{fig:PDF_CuCl_bond}
\end{figure}

Figure \ref{fig:PDF_CuCl_bond} shows the Cu--Cl bond distances obtained when fitting the room temperature Jahn-Teller distorted structure to the 100, 300, 400, 450 and 500~K datasets over different ranges of $r$. The 3 unique bonds of the P6$_1$22 model are plotted (each bond occurs twice per CuCl$_6$ octahedra for a total of 6 bonds). For all temperatures and all $r_\textrm{max}$, the fits indicate the presence of 4 short and 2 long bonds. In particular, the shortest Cu--Cl bond takes a constant value at all temperatures, close to the reported room temperature value of 2.281(4)~\AA.\cite{schlueter_redetermination_1966} This explains why the shortest Cu--Cl bond of the P6$_3$/mmc split-site structure (2.3300~\AA)\cite{crama_jahnteller_1981} poorly fits the PDF data. The second-shortest Cu--Cl bond gets longer as the temperature increases. At 450 and 500~K, it is longer ($\sim$2.4~\AA) than those reported by the room temperature P6$_1$22 structure (2.354(4)~\AA) and by the high temperature P6$_3$/mmc structure (2.3606~\AA). The P6$_3$/mmc split-site structure also slightly overestimates the longest Cu--Cl bond (2.7221~\AA~instead of~$\lesssim$2.7~\AA). 

The robust Cu--Cl distances over short and long ranges serve to confirm that the Jahn-Teller distortions of the CuCl$_6$ octahedra persist at all temperatures. This is in agreement with the results of many other studies, such as x-ray diffraction, electron paramagnetic resonance and magnetic susceptibility studies, which previously concluded that the octahedra must be Jahn-Teller distorted in the high temperature phase.\cite{crama_jahnteller_1981, Tanaka_1981_EPR, tanaka_electron_1985, haije_magnetic_1986} 
Below the phase transition, the long axes of 6 symmetry-equivalent CuCl$_6$ octahedra form an ordered chiral motif along the $c$-axis. Above the phase transition, the P6$_3$mc or P6$_3$/mmc structures are plausible average approximations because the orientations of the octahedral long axes become uncorrelated over long ranges. Octahedra may be uncorrelated from their immediate neighboring octahedra, or there may be ordered domains of few octahedra. Within those domains, the octahedra would be ordered, with the domains being orientationally disordered with respect to each other. To verify the existence of a short-range order and determine its length scale, we take a closer look at the refined atom positions. The weighted $R$ value of each fit is available in the Supporting Information, along with the refined values of the isotropic displacement parameters and delta1 or delta2. 

\begin{figure}[t]
    \centering
    \includegraphics[width=\columnwidth]{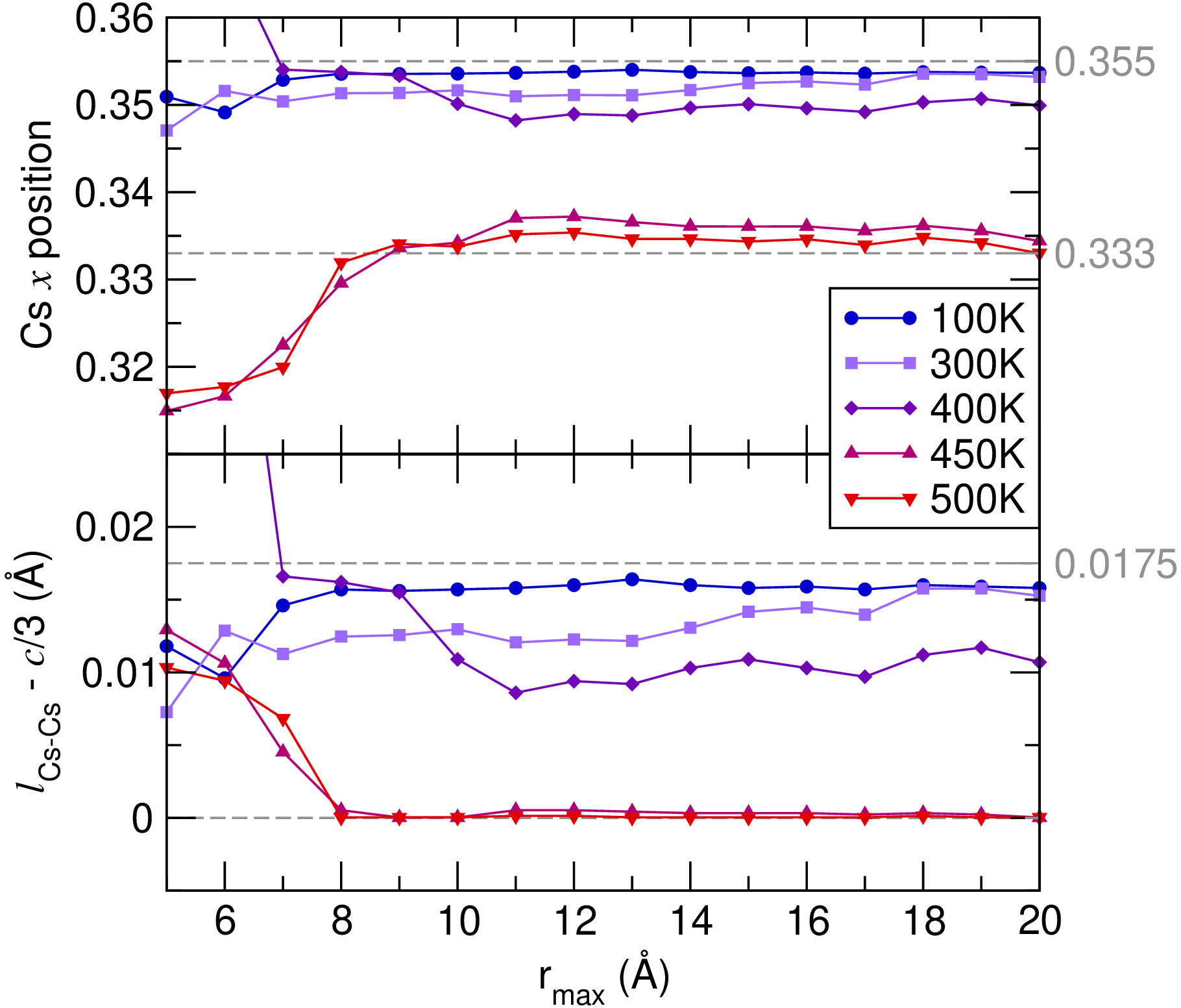}
    \caption{Cesium positional parameters obtained from fitting the P6$_1$22 structure on the 100~K (blue circles), 300~K (lilac squares), 400~K (purple diamonds), 450~K (burgundy up triangles) and 500~K (red down triangles) PDF datasets with varying $r_\textrm{max}$. Top: $x$ coordinate of the Cs atom at position ($x$, 2$x$, 1/4). Bottom: Difference between the out-of-plane Cs--Cs bond length and one third of the $c$-axis length. In both plots, grey dotted lines display the reported values for the P6$_1$22, P6$_3$mc and P6$_3$/mmc structures (see text for more details). At 100, 300 and 400~K, the $x$ coordinate is mostly constant. At 450 and 500~K, the value increases before it starts to plateau at around 8~\AA, a sign of short-range correlations. }
    \label{fig:PDF_Cs_position}
\end{figure}

\subsubsection{Cs coordinates}

In the P6$_1$22 structure, the standard coordinate of the Cs ion is ($x$, 2$x$, 1/4).\cite{schlueter_redetermination_1966} All other Cs coordinates can be obtained by symmetry. In each unit cell, 3 Cs ions are stacked along $c$. They have small in-plane displacements giving rise to the expected chiral motif (see Fig.~\ref{fig:Structures}(d)-(e)). The out-of-plane Cs--Cs distances (6.0767(4)~\AA) are then larger than 1/3 of the $c$-axis lattice parameter (6.0592~\AA).\cite{schlueter_redetermination_1966} This is in contrast to the achiral high temperature models, where symmetry constrains prevent any in-plane displacements of the Cs ions. Achiral positioning of the Cs ions would then result in out-of-plane Cs--Cs bond lengths of exactly $c$/3.
The upper and lower parts of Fig.~\ref{fig:PDF_Cs_position} show the refined value of $x$ and the difference between this out-of-plane Cs--Cs bond length and $c$/3 for different values of $r_\textrm{max}$. 

Fits for the 100, 300 and 400~K datasets result in almost perfectly constant $x$ coordinate close to the reported value of 0.35458.\cite{schlueter_redetermination_1966} The difference between the out-of-plane Cs--Cs bond length and $c$/3 is also constant and non-zero, as expected for the chiral phase. Two low $r$ outliers in the 400~K data are not shown in the figure and could be explained by the limited Cs bond information below 6~\AA~(the signal from the out-of-plane Cs--Cs bond is above 6~\AA).

Fits for the 450 and 500~K data show a very different trend than the low temperature data. The low $r_\textrm{max}$ results are very different from the mid to high $r_\textrm{max}$ results, a sign of short-range order in the high temperature phase.
For $r_\textrm{max}$ less than 7~\AA, $x$ is close to 0.315, then it increases and plateaus to about 0.335, which is close to the standard Cs coordinate (1/3, 2/3, 3/4) of both the P6$_3$mc and P6$_3$/mmc high-symmetry structures.\cite{kroese_high-temperature_1974,crama_jahnteller_1981} The plot of $l_\textrm{Cs--Cs} - c/3$ highlights that the change in coordinate occurs at $r_\textrm{max}=8$~\AA, quantifying the length scale of the short-range order. From $r_\textrm{max}=8$ to 20~\AA, $l_\textrm{Cs--Cs} - c/3$ is zero, indicating that the cesium atoms take achiral positions. Below $r_\textrm{max}=8$~\AA, the difference between the out-of-plane Cs--Cs bond length and $c$/3 is similar to that of the low temperature data, highlighting the resemblance of the local high temperature structure with the low temperature structure.

\begin{figure}
    \centering
    \includegraphics[width=\columnwidth]{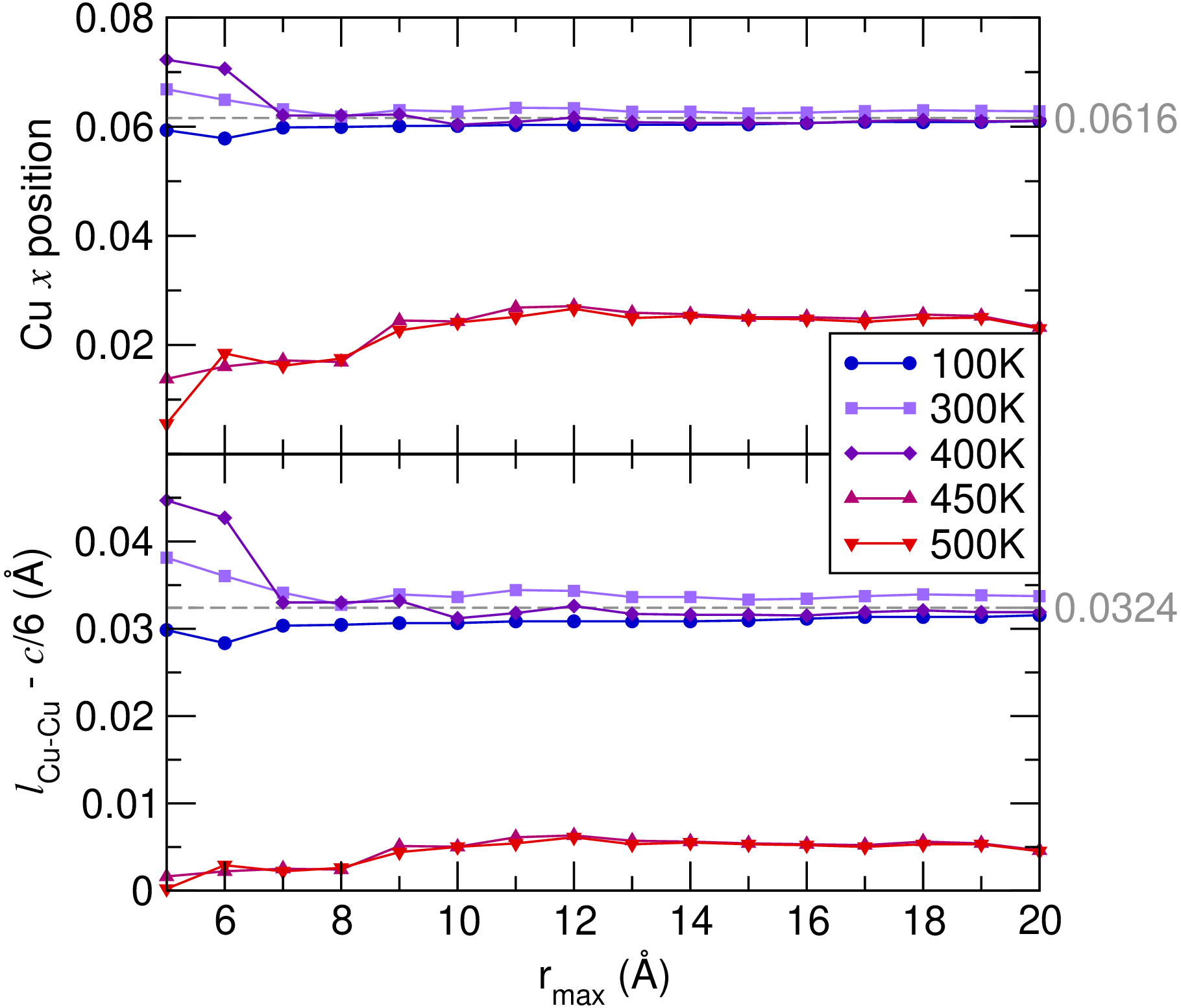}
    \caption{Copper positional parameters obtained from fitting the P6$_1$22 structure on the 100~K (blue), 300~K (lilac), 400~K (purple), 450~K (burgundy) and 500~K (red) PDF datasets with varying $r_\textrm{max}$. Top: $x$ coordinate of the Cu atom at position ($x$, 0, 0). Bottom: Difference between the out-of-plane Cu--Cu bond length and one sixth of the $c$-axis length. In both plots, grey dotted lines display the reported values for the P6$_1$22 structure. At 100, 300 and 400~K, the $x$ coordinate is mostly constant. As in Fig. \ref{fig:PDF_Cl_position}, the rise up to about 9~\AA\ indicates short-range correlations. }
    \label{fig:PDF_Cu_position}
\end{figure}

\subsubsection{Cu coordinates}

In the P6$_1$22 structure, the standard coordinate of the Cu ion is ($x$, 0, 0)\cite{schlueter_redetermination_1966} and all other Cu coordinates can be expressed in terms of $x$. The Cu atoms form spirals along the $c$-axis with 6 atoms per spiral per unit cell. The chiral motif induces a small in-plane displacement ($x\neq0$) that makes the out-of-plane Cu--Cu bond lengths (3.0620(4)~\AA) longer than one sixth of the $c$-axis lattice parameter (3.0296~\AA). In contrast, an achiral Cu position ($x=0$) would result in an out-of-plane Cu--Cu bond length of exactly $c$/6. The top and bottom portions of Fig.~\ref{fig:PDF_Cu_position} show the refined value of $x$ and the length difference between the out-of-plane Cu--Cu bond and $c$/6 for different values of $r_\textrm{max}$. For 100, 300 and 400~K, both plots show constant positions and bond lengths at all $r_\textrm{max}$, with $x$ close to the reported value of 0.06160.\cite{schlueter_redetermination_1966} 
The Cu ion positions and the out-of-plane Cu--Cu distances minus $c$/6 are then consistent with the expected low symmetry chiral structure. 
Fluctuations at $r_\textrm{max}=5$ and 6~\AA~can be explained by the limited Cu bonding information. 

At 450 and 500~K, the Cu $x$ value and the difference between the out-of-plane Cu--Cu bond length and $c$/6 are smaller than at low temperatures, indicating that the copper ions become more linear. The refined Cu positions are then closer to those reported for the high symmetry structures ($x=0$ for both P6$_3$mc and P6$_3$/mmc). Still, as seen in Fig.~\ref{fig:PDF_Cu_position}, the values of the copper ion position and of the bond length minus $c$/6 increase slightly as $r_\textrm{max}$ increases, up to about 9~\AA, at which point the values stabilize and tend to approximately 0.02 for $x$ and 0.005 for the bond length difference. These differences between the short-range structure and the long-range structure are an indication of short-range order above the phase transition temperature.

\subsubsection{Cl coordinates}

\begin{figure}
    \centering
    \includegraphics[width=\columnwidth]{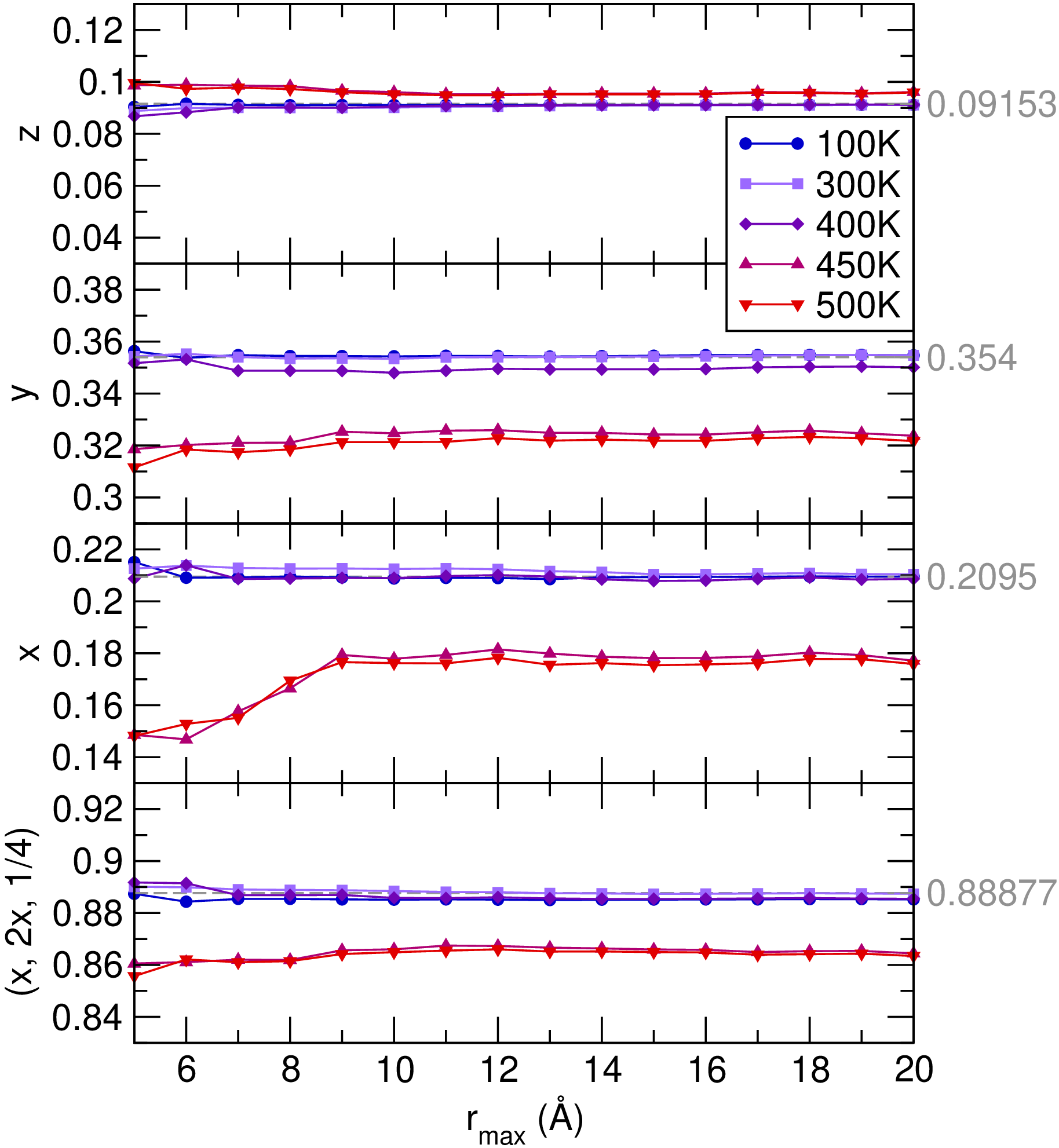}
    \caption{Coordinates of the Cl atoms as the P6$_1$22 structure is fitted to PDF datasets with different $r_\textrm{max}$ cutoff. The bottom panel displays the value of $x$ for the Wyckoff position 6b with coordinate ($x$, 2$x$, 1/4), while the three top panels display the values of $x$, $y$ and $z$ for the Wyckoff position 12c. Fitted values for datasets at 100, 300, 400, 450 and 500~K are shown with blue circles, lavender squares, purple diamonds, burgundy up triangles and red down triangles, respectively. In all plots, grey dotted lines (overlapping the low temperature data) display the reported values for the P6$_1$22 structure. At 450 and 500~K, the $x$ coordinate of the 12c site is most indicative of short-range correlations. }
    \label{fig:PDF_Cl_position}
\end{figure}

The Cl ions of structure P6$_1$22 are distributed over two different Wyckoff sites; 6b with standard coordinate ($x$, 2$x$, 1/4), and 12c with standard coordinate ($x$, $y$, $z$).\cite{schlueter_redetermination_1966} Figure~\ref{fig:PDF_Cl_position} shows the fitted values of these four coordinates for different values of $r_\textrm{max}$ and for the different temperature datasets. For 100, 300 and 400~K, all coordinates are constant as $r_\textrm{max}$ increases and are very similar to the values found in the literature. For the 6b site, the reported value of $x$ is 0.88770, and for the 12c site, the reported coordinate is (0.20950, 0.35400, 0.09153).\cite{schlueter_redetermination_1966} The low temperature data displays no difference between the short and long range structure, and is a good fit to the expected P6$_1$22 model. Multiple strong peaks for Cu--Cl and Cl--Cl bonds are present in the data for all ranges of $r_\textrm{max}$, so no fluctuations are seen in the low temperature small $r_\textrm{max}$ fits. 

At 450 and 500~K, all coordinates are constant except for the $x$ coordinate of the 12c site (second panel from the bottom in Fig.~\ref{fig:PDF_Cl_position}). The value is lower than at low temperatures and it slightly increases before becoming constant at $r_\textrm{max}=9$~\AA. This noticeable difference between the short- and long-range structure indicates the presence of short-range order at high temperatures.  

The Cl positions are significantly different at high and low temperatures, which explains the changes in Cu--Cl bond lengths seen in Fig.~\ref{fig:PDF_CuCl_bond}. 
Above the phase transition, the 6b Cl sites (bottom panel in Fig.~\ref{fig:PDF_Cl_position}) are further away from the nearest Cu atom, resulting in longer second-shortest Cu--Cl bonds in Fig.~\ref{fig:PDF_CuCl_bond}. 
At 450 and 500~K, the 12c Cl sites (top 3 panels in Fig.~\ref{fig:PDF_Cl_position}) get a little bit closer to the nearest Cu atom in such a way that the shortest Cu--Cl bond is minimally impacted, but the long Cu--Cl bond made with the next nearest Cu atom gets shorter (see Fig.~\ref{fig:PDF_CuCl_bond}). The temperature-dependent changes in Cu--Cl bond lengths are also affected by the Cu atom positions, which vary with temperature (see Fig.~\ref{fig:PDF_Cu_position}) and are correlated with the Cl atom positions. 

\subsubsection{Short-range correlation length}

At 450 and 500~K, the short and long-range fits give different values for the Cs, Cu and Cl ion positions, indicating the presence of short-range correlations. The change in values occurs around 8 or 9~\AA. This suggests that the Jahn-Teller distorted CuCl$_6$ octahedra are ordered over domains of length 8-9~\AA. It is interesting to note that this corresponds to a length of approximately 3 CuCl$_6$ octahedra (9.186~\AA). 

The short-range ordering that we have uncovered is subtle and was likely missed by previous studies due to the strong long-range randomness of the structure. 
From 1986 to 1989, experimental and theoretical works by Graf, Schotte and their co-authors\cite{graf_jahn-teller_1986, schotte_theory_1987, graf_quasi-elastic_1989, schotte_elastic_1989} investigated the diffuse scattering observed in neutron diffraction data. They claimed that there were no signs of intrachain short-range correlations in the high temperature structure of CsCuCl$_3$, and attributed the shape of the diffuse scattering to the presence of nonperiodic changes to the structure factor,  which they modeled using dense impurity Huang scattering. In 1987, Schotte calculated the diffuse scattering produced by the random orientation of elongated CuCl$_6$ octahedra with and without corrections for nearest-neighbor correlations. Upon comparison with experimental data, they concluded that there are no signs of short-range correlations in specific crystallographic directions.\cite{schotte_theory_1987} 
Even though their later work pointed out an error in the scaling factor for the uncorrelated scattering cross section,\cite{schotte_elastic_1989} the relative intensity arising from short-range ordering compared to the long-range disorder was not reinvestigated. We agree that the data from these neutron studies is likely dominated by the randomness of the long-range structure, but the data need to be quantitatively reexamined with the corrected model to determine how much intensity arises from short-range correlations.

Pair distribution function analysis is well-suited to investigate materials having local structures that are different from their average structures.\cite{qiu_orbital_2005,dey_monoclinic_2022,nagle-cocco_displacive_2024,shoemaker_unraveling_2009,Shoemaker2010,jiang_probing_2021} In this study, we opted to use small box least squares fitting to investigate correlation lengths in CsCuCl$_3$. Our results show clear evidence of short-range correlations on the order of 8 to 9~\AA. It would be interesting to compare our results with those of large box reverse Monte Carlo modeling and to perform additional high temperature measurements to help identify a potential temperature dependence of the short-range correlation length.

\subsection{Stability of CsCuCl$_3$ by diffraction and calorimetry}\label{sec:stability}

While the presence of a structural transition at 423~K in CsCuCl$_3$ is widely accepted, other unspecified transitions have been proposed in previous studies. For example, Fernández \textit{et al.}\ report additional transitions at 511 and 535~K based on small peaks in differential thermal analysis.\cite{fernandez_new_1976} Vasudevan \textit{et al.}\ and Bázan \textit{et al.}\ also observed these transitions by far infrared spectroscopy and differential thermal analysis, respectively.\cite{VASUDEVAN197944, bazan_reduction_2011} We believe these transitions are extrinsic in origin.

\begin{figure}
    \centering
    \includegraphics[width=\columnwidth]{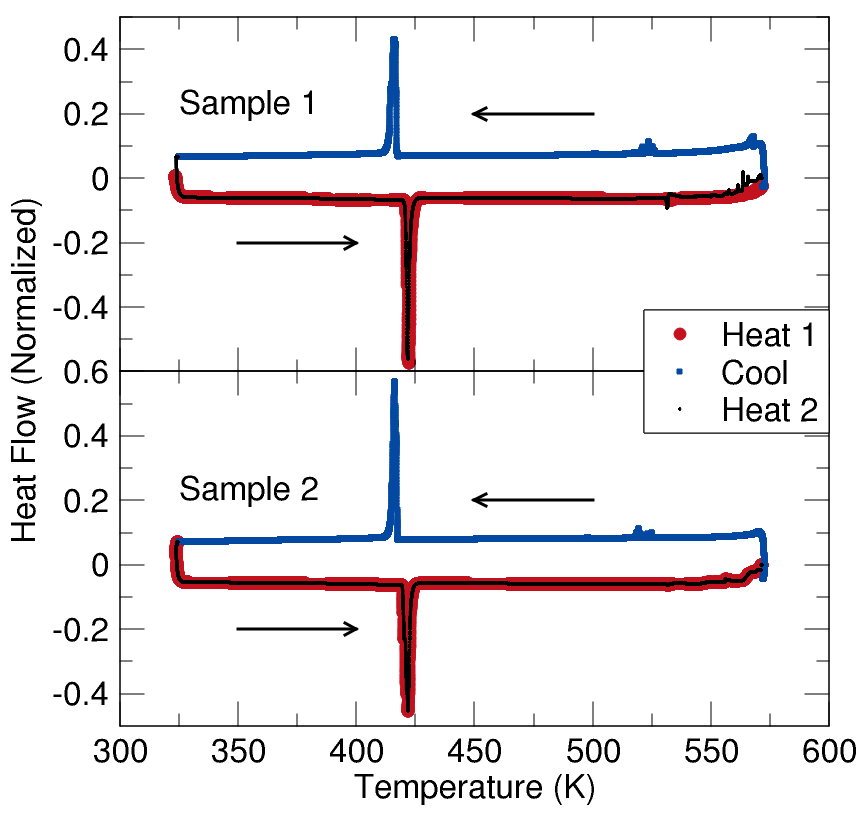}
    \caption{Differential scanning calorimetry (DSC) data collected on two samples of CsCuCl$_3$. The samples were heated (red thick line), cooled (blue line), and heated again (black thin line). Arrows indicate the direction of heating/cooling. The structural phase transition occurred at 422~K on heating and at 416~K on cooling.}
    \label{fig:DSC}
\end{figure}

To investigate the phase transitions of CsCuCl$_3$, we collected differential scanning calorimetry (DSC) data on two different samples, twice on heating and once on cooling; and we performed an \textit{in situ} powder x-ray diffraction (PXRD) experiment. 
Our DSC data (see Fig.~\ref{fig:DSC}) shows a phase transition at 422~K on heating and 416~K on cooling, which is in good agreement with the reported value of 423~K for the first-order chiral-achiral structural phase transition. Having some amount of thermal hysteresis is also in agreement with previous reports.\cite{fernandez_new_1976, hirotsu_optical_1975, Natarajan1971_phase_transition} The normalized enthalpies of the transitions are all between 1.39 and 1.74~kJ/mol.

\begin{figure}
    \centering
    \includegraphics[width=\columnwidth]{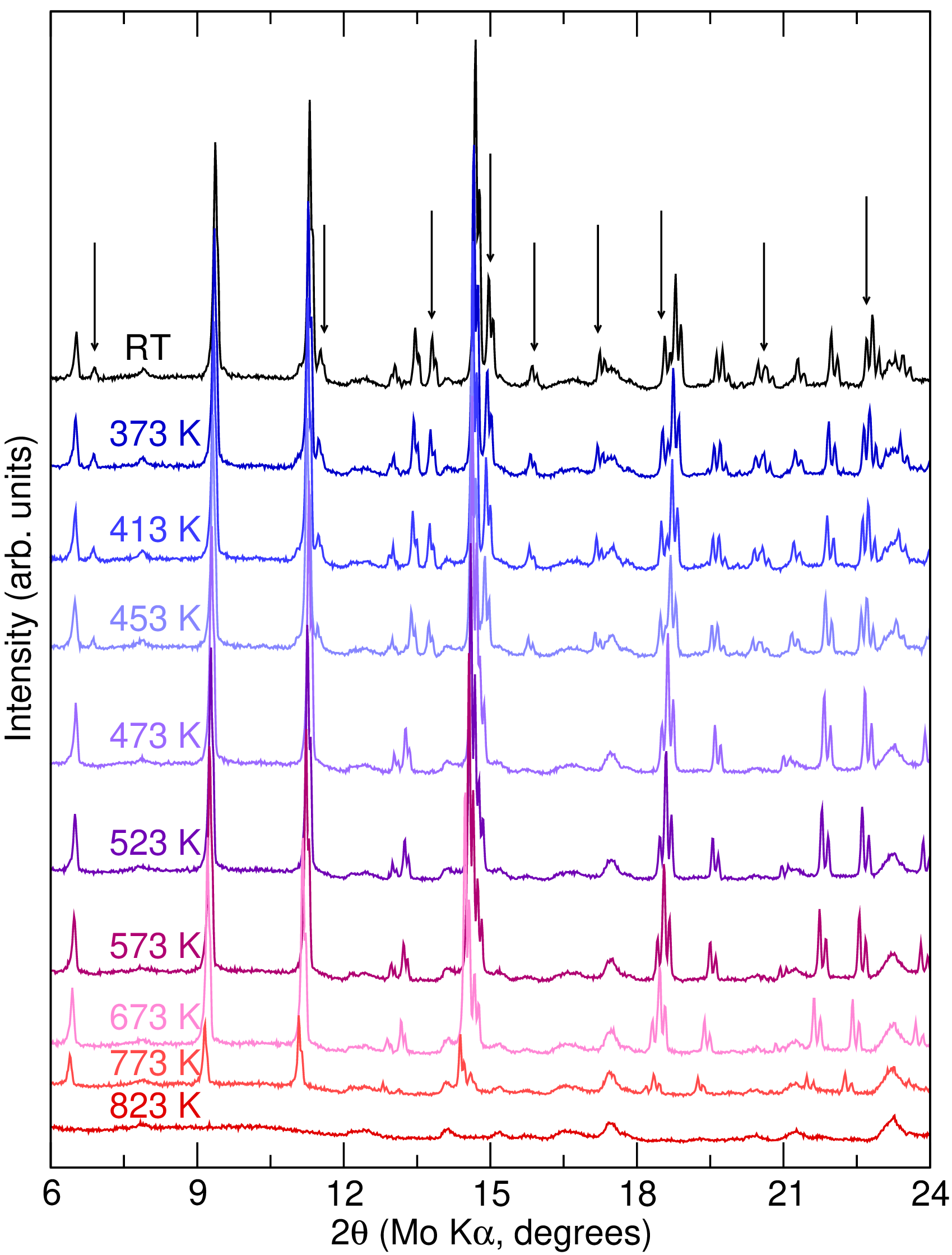}
    \caption{\textit{In situ} powder x-ray diffraction of CsCuCl$_3$ collected on heating. Because of furnace insulation effects, the structural phase transition happens between 453 and 473~K, instead of 423~K. Subtle changes in the peak profiles and intensities are signs of the phase transition; the relevant peaks are marked by black arrows in the room temperature data. No other structural changes are observed up to the melting of the compound. Rietveld refinements to some of the datasets are available in Fig. S9 of the Supporting Information. }
    \label{fig:insituPXRD}
\end{figure}

We did not observe transitions at 511 and 535~K in our DSC measurements (see Fig.~\ref{fig:DSC}) and we observed no structural changes above 473~K in our \textit{in situ} powder x-ray diffraction (PXRD) data (see Fig.~\ref{fig:insituPXRD}). 
Instead, our DSC datasets show a tiny noise-like signal on cooling around 523~K, which is the only signal (apart from the 423~K transition) reproduced in both samples. The signal is comprised of a few small peaks between 516 and 527~K with total normalized enthalpies of 0.12 and 0.08 kJ/mol.
All of the small signals in our datasets are similar in shape and in amplitude to the noise seen as the heating cycles end and the instrument switches to cooling mode. 

We attribute the signals at 511 and 535~K seen in other studies, and those at 523~K in our DSC data, to liquid inclusions (which can be obvious in optical microscopy or polishing) or minor phase impurities. These inclusions likely explain why  some crystals were seen to ``jump'' out of small crucibles or move laterally when heated. Trapped liquids may react with the crystals or containers to produce minor phases. 
We do not believe that CsCuCl$_3$ has any intrinsic phase transitions above the structural change at 423~K. 

\begin{figure}
    \centering
    \includegraphics[width=\columnwidth]{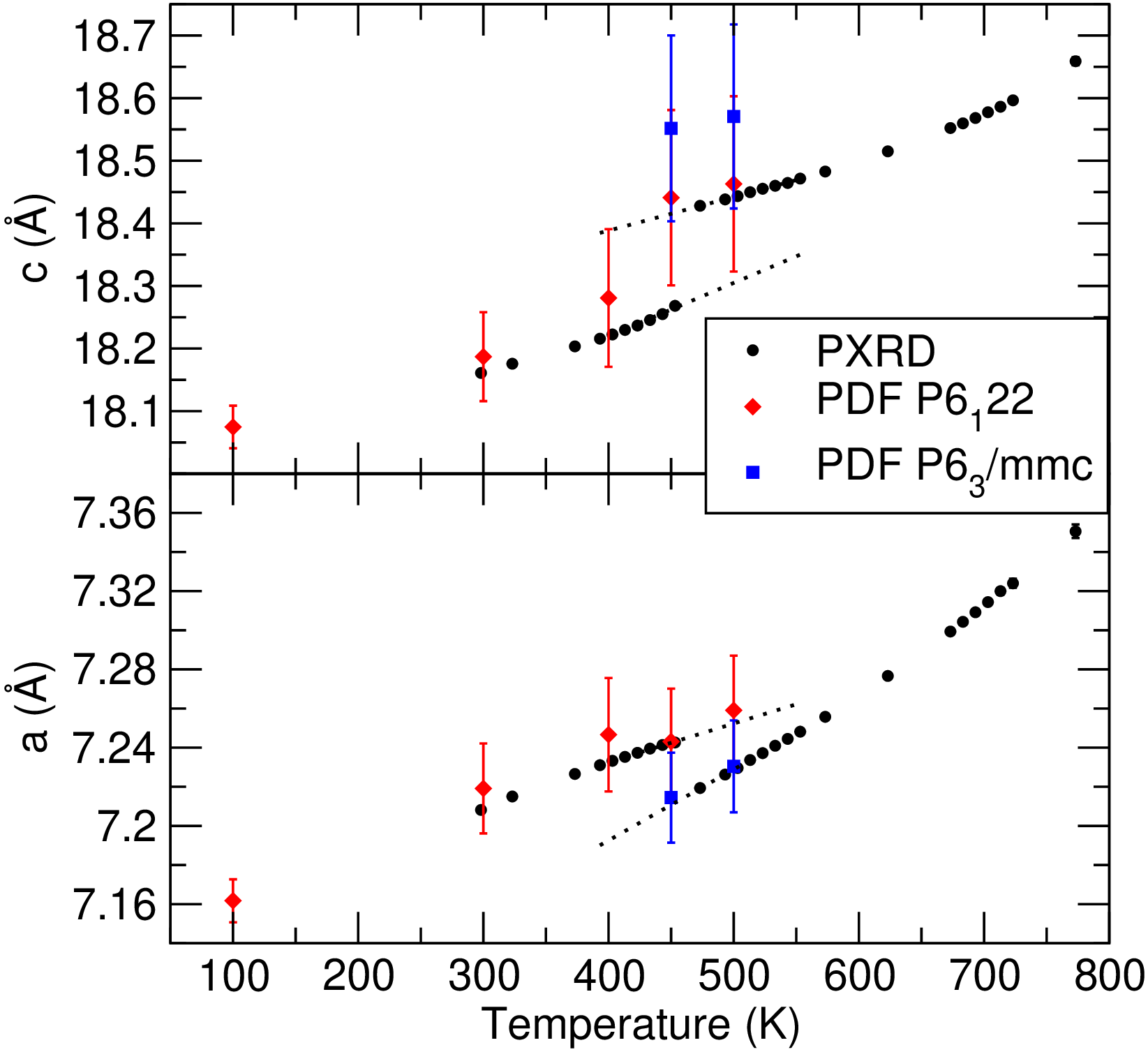}
    \caption{Lattice parameters of CsCuCl$_3$ at different temperatures. Black circles indicate data obtained from fitting powder x-ray diffraction data with the P6$_1$22 (298 to 453~K) and P6$_3$/mmc (473 to 773~K) structures. Red diamonds and blue squares indicate data obtained from fitting the P6$_1$22 or P6$_3$/mmc structures, respectively, to pair distribution function data. Error bars are plotted, but are sometimes too small to be seen. Black dotted lines extrapolate the PXRD data between 393 and 553~K to highlight the change in the lattice parameters (0.4\% contraction in $a$ and 0.8\% expansion in $c$). }
    \label{fig:latparam}
\end{figure}

PXRD data collected on heating is shown in Fig.~\ref{fig:insituPXRD} for selected temperatures. A zoomed-in view of the 6 to 24$^{\circ}$ 2$\theta$ range is presented to better highlight the subtle visual changes in the signal associated with the structural phase transition.
Black arrows above the room temperature data mark peaks 101, 111, 202, 106, 204, 205, 213, 207, 215, and 221, which disappear or merge with nearby peaks upon the phase transition (207 and 215 nearly overlap and share an arrow).
These changes can be observed between the data at 453 and 473~K. A larger version of Fig.\ref{fig:insituPXRD} is available in the Supporting Information.

The PXRD data shows that the chiral structure persisted up to 453~K, which is above the reported transition temperature of 423~K and the transition temperature we obtained from our DSC data. The discrepancy between these values can be attributed to insulation differences, with the DSC data giving a more precise and reliable value for the transition temperature. At 473~K, the PXRD data fits well to the high temperature achiral structure. No other structural changes are observed up to 773~K, where a decrease in the signal indicates that the sample is starting to melt. Lack of signal at 823~K indicates that the sample is completely melted. Comparison of data collected with and without the furnace indicate that minimal scattering from the furnace is present, which is visible in the 823~K dataset. On cooling (see the Supporting Information), the signal reappears around 723~K and increases in strength until about 573~K. After 573~K, further cooling does not increase the PXRD signal intensity. The room temperature PXRD data obtained after heating seems to indicate that the \textit{in situ} sample is preferentially oriented CsCuCl$_3$. 

To confirm that CsCuCl$_3$ is congruently melting, crystals from a different synthesis batch were ground to powder, sealed in an evacuated quartz tube, heated to 823~K at 10~K/min, held at 823\C for 30 min, and slow cooled at 2.5~K/min. This annealing melted the sample and PXRD confirmed that the sample was still CsCuCl$_3$. BASF (BAtch Scale Factor) parameters obtained from single crystal x-ray diffraction on 3 fragments indicated that the sample was of mixed chirality after melting. Soboleva \textit{et al.}\ reported that CsCuCl$_3$ melts and partially decomposes at 728~K.\cite{soboleva_investigation_1976} The reported melting temperature is in reasonable agreement with our \textit{in situ} PXRD results. However, we did not observe any partial decomposition and found that CsCuCl$_3$ is congruently melting. 

The PXRD data from 298 to 453~K was fit to the P6$_1$22 structure, while the PXRD data from 473 to 773~K was fit to the P6$_3$/mmc structure. 
The lattice parameters were refined, while the atom positions were fixed to the reported values. 
Fits of the data at 298, 473 and 673~K are available in the Supporting Information. Values for the lattice parameters are plotted in Fig.~\ref{fig:latparam} along with the values obtained from fits of the PDF data. Since the low temperature P6$_1$22 structure has a $c$-axis that is about 3 times longer than that of the high temperature P6$_3$/mmc structure, the refined value of the high temperature $c$-axis is tripled in Fig.~\ref{fig:latparam}. For the PDF data at 450 and 500~K, the lattice parameters obtained from fits of the P6$_1$22 and P6$_3$/mmc structures are present (for fit details, see the description of the P6$_1$22 r$_\textrm{max}$ fits and the description of the P6$_3$/mmc supercell fit in the previous sections). Figure~\ref{fig:latparam} shows that there is a noticeable step change in the lattice parameters, consistent with the first-order nature of the structural phase transition. This change is at the correct temperature (423~K) for the PDF data, but a temperature lag in the PXRD data artificially shifts the transition to a higher temperature. Black dotted lines show lattice parameter values extrapolated from the PXRD data between 393 and 453~K and between 473 and 553~K. These extrapolations indicate that the lattice undergoes a 0.4\% contraction along $a$ and a 0.8\% expansion along $c$ during the phase transition. Our results are in reasonable agreement with the dilatometry measurements of Hirotsu (1975).\cite{hirotsu_optical_1975}

\subsection{Absolute structure of pristine and heated CsCuCl$_3$}

Many studies have reported that CsCuCl$_3$ crystals obtained by aqueous precipitation are heterochiral. Measurements of optical activity\cite{hirotsu_optical_1975}, single crystal x-ray diffraction (SCXRD) Flack parameters\cite{kousaka_crystallographic_2009, kousaka_crystal_2014, inomata_temperature-dependent_2025, inomata_solvent_2026}, anomalous x-ray scattering\cite{koiso_determination_1996}, and resonant circularly polarized x-ray diffraction\cite{kousaka_crystallographic_2009, ohsumi_threedimensional_2013} have shown that as grown crystals had domains of different chirality. Methods to obtain homochiral crystals have been found; stirring\cite{kousaka_crystal_2014}, heating\cite{inomata_temperature-dependent_2025} and adding certain organic solvents\cite{inomata_solvent_2026} to the solution lead to the precipitation of homochiral crystals (based on measurements of SCXRD Flack parameters). Homochiral seeds can then be used to obtain larger homochiral crystals.\cite{kousaka_monochiral_2017}

\textit{In situ} measurements of optical rotation\cite{hirotsu_optical_1975} also showed that the rotation angle, initially non-zero, becomes null after heating and cooling CsCuCl$_3$ through the phase transition. This result seemed to indicate that fine enantiomorphic domains smaller than the probed region (1~mm) are forming as the low temperature structure is recovered.\cite{hirotsu_optical_1975}

\begin{figure}
    \centering
    \includegraphics[width=\columnwidth]{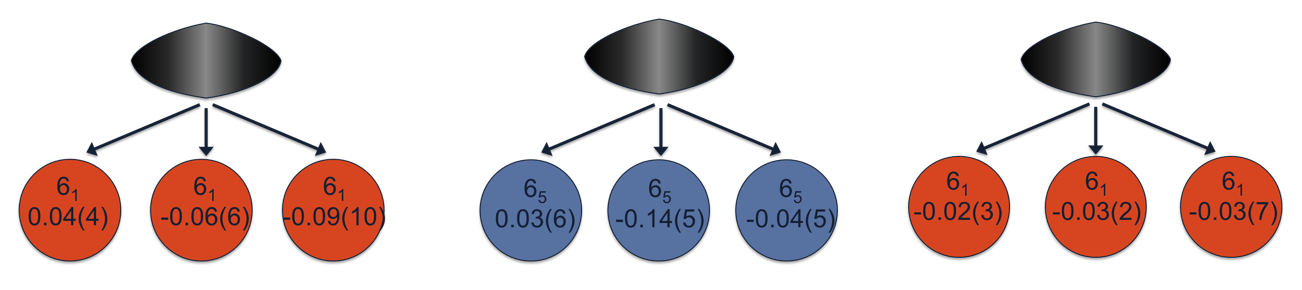}
    \caption{Space groups and Flack parameters obtained from single crystal x-ray diffraction of pristine CsCuCl$_3$ crystals. Three fragments (colored disks) are measured from each of the three crystals (black rhombuses). Space groups P6$_1$22 (red) and P6$_5$22 (blue) are abbreviated by 6$_1$ and 6$_5$, respectively. Near zero Flack parameters indicate that all fragments are homochiral. Fragments taken from the same crystal have the same handedness, suggesting that the crystals are homochiral. }
    \label{fig:RT_Flack}
\end{figure}

To verify the absolute structure of our as grown CsCuCl$_3$, we measured the SCXRD Flack parameters of a few crystals from two different precipitation batches. Figure~\ref{fig:RT_Flack} shows the Flack parameter and space group of 9 fragments of pristine CsCuCl$_3$ taken from 3 different crystals. Fragments had lengths from 100 to 300 microns with widths and thicknesses of a few tens of microns. Near zero Flack parameters indicate that all fragments are homochiral. Fragments obtained from the same crystal also display the same handedness. This suggests that the CsCuCl$_3$ crystals in our study are homochiral, in contrast to the many studies reporting heterochiral as grown crystals of CsCuCl$_3$.\cite{hirotsu_optical_1975, kousaka_crystallographic_2009, kousaka_crystal_2014, inomata_temperature-dependent_2025, inomata_solvent_2026, koiso_determination_1996, kousaka_crystallographic_2009, ohsumi_threedimensional_2013} 

\begin{figure}
    \centering
    \includegraphics[width=\columnwidth]{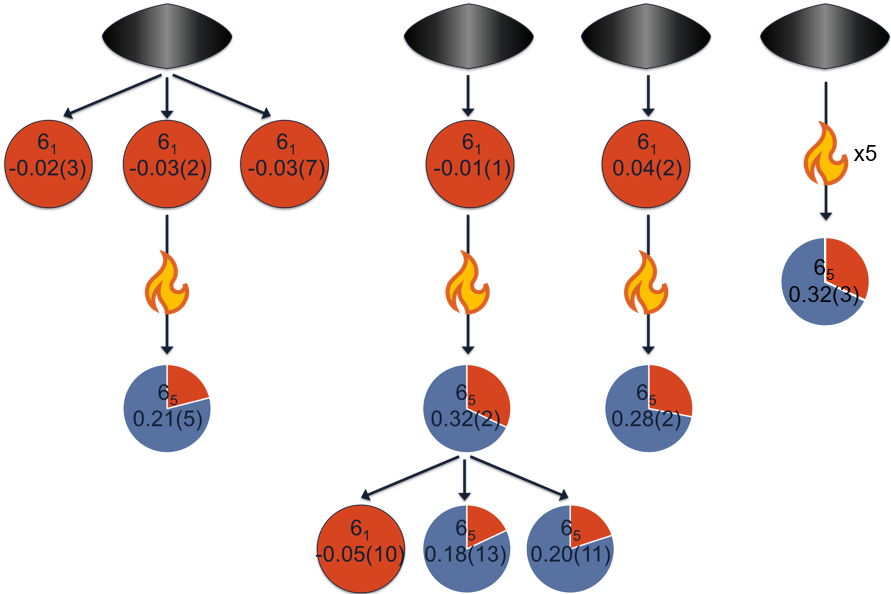}
    \caption{Space groups and Flack parameters obtained from \textit{in situ} and \textit{ex situ} single crystal x-ray diffraction of fragments (colored pie charts) of CsCuCl$_3$ crystals (black rhombuses). Space groups P6$_1$22 (red) and P6$_5$22 (blue) are abbreviated by 6$_1$ and 6$_5$, respectively. Crystal fragments have near zero Flack parameters before being heated to 450~K, indicating that they are homochiral at the start. After being heated and cooled through the phase transition (fire icon), the fragments are of mixed chirality. The last schematic to the right shows that a fragment taken from a crystal cycled 5 times through the phase transition is also heterochiral. }
    \label{fig:HT_Flack}
\end{figure}

To verify the effects of the phase transition on the chirality of CsCuCl$_3$, we measured the Flack parameters of 3 fragments before and after heating to 450~K and cooling back to room temperature. As seen in Fig.~\ref{fig:HT_Flack}, all 3 fragments, originally homochiral, are of mixed handedness after undergoing the phase transition. The dominant handedness is also changed by the heating and cooling. One heated fragment was broken in 3 pieces and the Flack parameter was measured again. Two of these smaller pieces are of mixed chirality and have an opposite dominant handedness compared to the third piece. The different results obtained for these 3 smaller pieces highlight again that the phase transition induces chiral domains. This result is also robust when we heat larger crystals, and not just small fragments. One crystal was heated and cooled through the phase transition 5 times before we measured the Flack parameter of a fragment. The measurement indicates that the crystal fragment is of mixed handedness, in contrast to the results of the pristine crystal fragments (see Fig.~\ref{fig:RT_Flack}). Larger crystals are seemingly unaffected by repeated temperature cycling and maintain their integrity and crystallinity. Our \textit{in situ} single crystal XRD experiments are then in agreement with the results of the \textit{in situ} optical rotation experiments by Hirotsu\cite{hirotsu_optical_1975}.

\section{Conclusions}

We reinvestigated the high temperature structure of CsCuCl$_3$ using pair distribution function analysis of neutron diffraction data. Our results show that the high temperature phase is Jahn-Teller distorted and that short-range correlations exist on the order of 8-9~\AA. To the best of our knowledge, this is the first evidence and characterization of the short-range order in the high temperature structure of CsCuCl$_3$. Additional high temperature measurements could help identify a potential temperature dependence of the short-range correlation length.

We used differential scanning calorimetry, \textit{in situ} powder x-ray diffraction, and annealing experiments to investigate the thermal stability of CsCuCl$_3$. We found no additional structural phase transitions and confirmed that CsCuCl$_3$ is congruently melting. The phase transition at 423~K leads to a 0.4\% contraction of the unit cell along $a$ and a 0.8\% expansion along $c$. 

Our single crystal x-ray diffraction experiments revealed that our as grown crystals are most likely homochiral, a surprising result in light of the many reports of as grown heterochiral crystals. 
Our results highlight that more research is needed to better understand the synthesis conditions leading to the formation of homochiral and heterochiral crystals of CsCuCl$_3$. 
More importantly, our experiments showed that the phase transition induces chiral domains in CsCuCl$_3$. This opens the door to new methods of controlling chirality in polymorphic materials. A more detailed investigation of the impact of the phase transition on domain formation in CsCuCl$_3$ is underway. Results of our \textit{in situ} chirality maps obtained by resonant circularly polarized x-ray diffraction will be available in an upcoming publication.

\section*{Supporting Information}

Additional synthesis details, powder XRD refinements, SEM-EDS images, and pair distribution function analysis.

\begin{acknowledgments}

This work was supported by the US Department of Energy (DOE), Basic Energy Science,
under award No. DE-SC0025305. 
Preliminary work on this compound was supported by the Center for Quantum Sensing and Quantum Materials, an Energy Frontier Research Center funded by the U.S. Department of Energy, Office of Science, Basic Energy Sciences, under Award DE-SC0021238.
Neutron scattering was performed at the Spallation Neutron Source (IPTS-36285), a DOE Office of Science User Facility operated by the Oak Ridge National Laboratory. Synthesis and characterization were carried out in the Materials Research Laboratory Central Research Facilities, University of Illinois. 
EAP was partially funded by a Ludo Frevel Crystallography Scholarship awarded by the International Centre for Diffraction Data. EAP would like to thank Jack D'Amelio for useful discussions.

\end{acknowledgments}

\bibliographystyle{unsrt}

\bibliography{CsCuCl3.bib}

@article{Farrow_2007,
doi = {10.1088/0953-8984/19/33/335219},
url = {https://doi.org/10.1088/0953-8984/19/33/335219},
year = {2007},
publisher = {},
volume = {19},
number = {33},
pages = {335219},
author = {Farrow, C. L. and Juhas, P. and Liu, J. W. and Bryndin, D. and Božin, E. S. and Bloch, J. and Proffen, Th. and Billinge, S. J. L.},
title = {{PDFfit2} and {PDFgui}: computer programs for studying nanostructure in crystals},
journal = {Journal of Physics: Condensed Matter}
}

@article{GSAS2,
author = "Toby, B. H. and Von Dreele, R. B.",
title = "{{\it GSAS-II}: the genesis of a modern open-source all purpose crystallography software package}",
journal = "Journal of Applied Crystallography",
year = "2013",
volume = "46",
number = "2",
pages = "544--549",
doi = {10.1107/S0021889813003531},
url = {https://doi.org/10.1107/S0021889813003531},
}

@article{Momma:db5098,
author = "Momma, K. and Izumi, F.",
title = "{{\it VESTA3} for three-dimensional visualization of crystal, volumetric and morphology data}",
journal = "Journal of Applied Crystallography",
year = "2011",
volume = "44",
number = "6",
pages = "1272--1276",
doi = {10.1107/S0021889811038970},
url = {https://doi.org/10.1107/S0021889811038970},
}

@manual{apex3,
  title        = {APEX3},
  author       = {{Bruker}},
  organization = {Bruker AXS Inc.},
  address      = {Madison, Wisconsin, USA},
  year         = {2014}
}

@article{krause_comparison_2015,
	title = {Comparison of silver and molybdenum microfocus {X}-ray sources for single-crystal structure determination},
	volume = {48},
	issn = {0021-8898},
	doi = {10.1107/S1600576714022985},
	language = {eng},
	number = {Pt 1},
	journal = {Journal of Applied Crystallography},
	author = {Krause, L. and Herbst-Irmer, R. and Sheldrick, G. M. and Stalke, D.},
	year = {2015},
	pages = {3--10},
}

@article{sheldrick_crystal_2015C,
	title = {Crystal structure refinement with {SHELXL}},
	volume = {71},
	issn = {2053-2296},
	doi = {10.1107/S2053229614024218},
	language = {eng},
	number = {Pt 1},
	journal = {Acta Crystallographica. Section C, Structural Chemistry},
	author = {Sheldrick, G. M.},
	year = {2015},
	pages = {3--8},
}

@article{sheldrick_shelxt_2015A,
	title = {\textit{{SHELXT}} – {Integrated} space-group and crystal-structure determination},
	volume = {71},
	issn = {2053-2733},
	url = {https://journals.iucr.org/paper?S2053273314026370},
	doi = {10.1107/S2053273314026370},
	language = {en},
	number = {1},
	urldate = {2026-05-18},
	journal = {Acta Crystallographica Section A Foundations and Advances},
	author = {Sheldrick, G. M.},
	year = {2015},
	pages = {3--8},
}

@article{NOMAD,
title = "The {Nanoscale} {Ordered} {MAterials} {Diffractometer} {NOMAD} at the {Spallation} {Neutron} {Source} {SNS}",
author = "J. Neuefeind and M. Feygenson and J. Carruth and R. Hoffmann and Chipley, K. K.",
year = "2012",
day = "15",
doi = "10.1016/j.nimb.2012.05.037",
language = "English",
volume = "287",
pages = "68--75",
journal = "Nuclear Instruments and Methods in Physics Research, Section B: Beam Interactions with Materials and Atoms",
issn = "0168-583X",
publisher = "Elsevier",
}

@article{ORNL,
    author = {Calder, S. and An, K. and Boehler, R. and Dela Cruz, C. R. and Frontzek, M. D. and Guthrie, M. and Haberl, B. and Huq, A. and Kimber, S. A. J. and Liu, J. and Molaison, J. J. and Neuefeind, J. and Page, K. and dos Santos, A. M. and Taddei, K. M. and Tulk, C. and Tucker, M. G.},
    title = {A suite-level review of the neutron powder diffraction instruments at {Oak} {Ridge} {National} {Laboratory}},
    journal = {Review of Scientific Instruments},
    volume = {89},
    number = {9},
    pages = {092701},
    year = {2018},
    issn = {0034-6748},
    doi = {10.1063/1.5033906},
    url = {https://doi.org/10.1063/1.5033906},
    eprint = {https://pubs.aip.org/aip/rsi/article-pdf/doi/10.1063/1.5033906/16758224/092701_1_online.pdf},
}

@article{Shoemaker2010,
  title = {Atomic displacements in the charge ice pyrochlore {Bi$_2$Ti$_2$O$_6$O}$^{\ensuremath{'}}$ studied by neutron total scattering},
  author = {Shoemaker, D. P. and Seshadri, R. and Hector, A. L. and Llobet, A. and Proffen, Th. and Fennie, C. J.},
  journal = {Physical Review B},
  volume = {81},
  issue = {14},
  pages = {144113},
  numpages = {9},
  year = {2010},
  publisher = {American Physical Society},
  doi = {10.1103/PhysRevB.81.144113},
  url = {https://link.aps.org/doi/10.1103/PhysRevB.81.144113}
}

@article{inomata_solvent_2026,
	title = {Solvent and {Achiral} {Crystalline} {Phase}-{Induced} {Chiral} {Resolution} of {CsCuCl$_3$}},
	volume = {26},
	issn = {1528-7483},
	url = {https://doi.org/10.1021/acs.cgd.5c01350},
	doi = {10.1021/acs.cgd.5c01350},
	number = {1},
	urldate = {2026-02-19},
	journal = {Crystal Growth \& Design},
	publisher = {American Chemical Society},
	author = {Inomata, Y. and Yamada, S. and Kida, T.},
	year = {2026},
	pages = {401--407},
}

@article{inomata_temperature-dependent_2025,
	title = {Temperature-{Dependent} {Spontaneous} {Resolution} of {CsCuCl}$_{\textrm{3}}$},
	copyright = {https://doi.org/10.15223/policy-029},
	issn = {1528-7483, 1528-7505},
	url = {https://pubs.acs.org/doi/10.1021/acs.cgd.5c00561},
	doi = {10.1021/acs.cgd.5c00561},
	language = {en},
	urldate = {2025-08-14},
	journal = {Crystal Growth \& Design},
    volume = {25},
    issue = {17},
	author = {Inomata, Y.},
	year = {2025},
	pages = {7081-7084},
}

@article{koiso_determination_1996,
	title = {Determination of the chiral structure of using anomalous x-ray scattering near the {Cs} {K} absorption edge},
	volume = {8},
	issn = {0953-8984},
	url = {https://doi.org/10.1088/0953-8984/8/38/010},
	doi = {10.1088/0953-8984/8/38/010},
	language = {en},
	number = {38},
	urldate = {2025-10-07},
	journal = {Journal of Physics: Condensed Matter},
	author = {Koiso, T. and Yamamoto, K. and Hata, Y. and Takahashi, Y. and Kita, E. and Ohshima, K. and Okamura, F. P.},
	year = {1996},
	pages = {7059},
}

@article{kousaka_monochiral_2017,
	title = {Monochiral helimagnetism in homochiral crystals of {CsCuCl$_3$}},
	volume = {1},
	url = {https://link.aps.org/doi/10.1103/PhysRevMaterials.1.071402},
	doi = {10.1103/PhysRevMaterials.1.071402},
	number = {7},
	urldate = {2025-08-14},
	journal = {Physical Review Materials},
	publisher = {American Physical Society},
	author = {Kousaka, Y. and Koyama, T. and Ohishi, K. and Kakurai, K. and Hutanu, V. and Ohsumi, H. and Arima, T. and Tokuda, A. and Suzuki, M. and Kawamura, N. and Nakao, A. and Hanashima, T. and Suzuki, J. and Campo, J. and Miyamoto, Y. and Sera, A. and Inoue, K. and Akimitsu, J.},
	year = {2017},
	pages = {071402},
}

@article{kousaka_crystallographic_2009,
	title = {Crystallographic {Chirality} of {CsCuCl}$_{\textrm{3}}$ {Probed} by {Resonant} {Circularly}-{Polarized} {Hard} {X}-ray {Diffraction}},
	volume = {78},
	issn = {0031-9015, 1347-4073},
	url = {http://journals.jps.jp/doi/10.1143/JPSJ.78.123601},
	doi = {10.1143/JPSJ.78.123601},
	language = {en},
	number = {12},
	urldate = {2025-08-14},
	journal = {Journal of the Physical Society of Japan},
	author = {Kousaka, Y. and Ohsumi, H. and Komesu, T. and Arima, T. and Takata, M. and Sakai, S. and Akita, M. and Inoue, K. and Yokobori, T. and Nakao, Y. and Kaya, E. and Akimitsu, J.},
	year = {2009},
	pages = {123601},
}

@article{ohsumi_threedimensional_2013,
	title = {Three‐{Dimensional} {Near}‐{Surface} {Imaging} of {Chirality} {Domains} with {Circularly} {Polarized} {X}‐rays},
	volume = {52},
	issn = {1433-7851, 1521-3773},
	url = {https://onlinelibrary.wiley.com/doi/10.1002/anie.201303023},
	doi = {10.1002/anie.201303023},
	language = {en},
	number = {33},
	urldate = {2025-08-14},
	journal = {Angewandte Chemie International Edition},
	author = {Ohsumi, H. and Tokuda, A. and Takeshita, S. and Takata, M. and Suzuki, M. and Kawamura, N. and Kousaka, Y. and Akimitsu, J. and Arima, T.},
	year = {2013},
	pages = {8718--8721},
}

@article{adachi_helical_1980,
	title = {Helical {Magnetic} {Structure} in {CsCuCl$_3$}},
	volume = {49},
	issn = {0031-9015},
	url = {https://journals.jps.jp/doi/10.1143/JPSJ.49.545},
	doi = {10.1143/JPSJ.49.545},
	number = {2},
	urldate = {2025-08-14},
	journal = {Journal of the Physical Society of Japan},
	publisher = {The Physical Society of Japan},
	author = {Adachi, K. and Achiwa, N. and Mekata, M.},
	year = {1980},
	pages = {545--553},
}

@article{maaskant_jahn-teller-induced_1986,
	title = {On the {Jahn}-{Teller}-induced helical deformations in {CsCuCl$_3$}},
	volume = {19},
	issn = {0022-3719},
	url = {https://doi.org/10.1088/0022-3719/19/27/007},
	doi = {10.1088/0022-3719/19/27/007},
	language = {en},
	number = {27},
	urldate = {2026-05-06},
	journal = {Journal of Physics C: Solid State Physics},
	author = {Maaskant, W. J. A. and Haije, W. G.},
	year = {1986},
	pages = {5295},
}

@article{tazuke_magnetic_1981,
	title = {Magnetic {Susceptibility} {Study} of {CsCuCl}$_{\textrm{3}}$},
	volume = {50},
	issn = {0031-9015, 1347-4073},
	url = {http://journals.jps.jp/doi/10.1143/JPSJ.50.3919},
	doi = {10.1143/JPSJ.50.3919},
	language = {en},
	number = {12},
	urldate = {2026-05-06},
	journal = {Journal of the Physical Society of Japan},
	author = {Tazuke, Y. and Tanaka, H. and Iio, K. and Nagata, K.},
	year = {1981},
	pages = {3919--3924},
}

@article{Tanaka_1981_EPR,
author = {Tanaka ,H. and Iio ,K. and Nagata ,K.},
title = {Influence of {Cooperative} {Jahn}-{Teller} {Effect} in {CsCuCl$_3$} {Crystals} on the {Broadening} of {EPR} {Lines}},
journal = {Journal of the Physical Society of Japan},
volume = {50},
number = {3},
pages = {727-728},
year = {1981},
doi = {10.1143/JPSJ.50.727},
URL = {https://doi.org/10.1143/JPSJ.50.727},
eprint = {https://doi.org/10.1143/JPSJ.50.727}
}

@article{Laiho1973_electrical_optical,
author = {Laiho, R. and Natarajan, M. and Kaira, M.},
title = {Some electrical and optical studies in {CsCuCl$_3$} crystals},
journal = {physica status solidi (a)},
volume = {15},
number = {1},
pages = {311-317},
doi = {https://doi.org/10.1002/pssa.2210150135},
url = {https://onlinelibrary.wiley.com/doi/abs/10.1002/pssa.2210150135},
eprint = {https://onlinelibrary.wiley.com/doi/pdf/10.1002/pssa.2210150135},
year = {1973}
}

@article{Natarajan1971_phase_transition,
author = {Natarajan, M. and Prakash, B.},
title = {Phase transitions in {ABX$_3$} type halides},
journal = {physica status solidi (a)},
volume = {4},
number = {3},
pages = {K167-K172},
doi = {https://doi.org/10.1002/pssa.2210040331},
url = {https://onlinelibrary.wiley.com/doi/abs/10.1002/pssa.2210040331},
eprint = {https://onlinelibrary.wiley.com/doi/pdf/10.1002/pssa.2210040331},
year = {1971}
}

@article{bazan_reduction_2011,
	title = {Reduction of {Cu}({II}) to {Cu}({I}) in solid {CsCuCl$_3$}},
	volume = {125},
	copyright = {https://www.elsevier.com/tdm/userlicense/1.0/},
	issn = {02540584},
	url = {https://linkinghub.elsevier.com/retrieve/pii/S0254058410008503},
	doi = {10.1016/j.matchemphys.2010.10.019},
	language = {en},
	number = {3},
	urldate = {2025-08-14},
	journal = {Materials Chemistry and Physics},
	author = {Bazán, J. C. and Lescano, G. M. and Prat, M. R. and Sagua, A.},
	year = {2011},
	pages = {542--547},
}

@article{hirotsu_optical_1975,
	title = {Some optical and thermal properties of {CsCuCl}$_{\textrm{3}}$ and its phase transition near {423K}},
	volume = {8},
	issn = {0022-3719},
	url = {https://iopscience.iop.org/article/10.1088/0022-3719/8/1/003},
	doi = {10.1088/0022-3719/8/1/003},
	language = {en},
	number = {1},
	urldate = {2025-08-14},
	journal = {Journal of Physics C: Solid State Physics},
	author = {Hirotsu, S.},
	year = {1975},
	pages = {L12--L16},
}

@article{hirotsu_jahn-teller_1977,
	title = {Jahn-{Teller} induced phase transition in {CsCuCl}$_{\textrm{3}}$ : structural phase transition with helical atomic displacements},
	volume = {10},
	issn = {0022-3719},
	shorttitle = {Jahn-{Teller} induced phase transition in {CsCuCl}$_{\textrm{3}}$},
	url = {https://iopscience.iop.org/article/10.1088/0022-3719/10/7/008},
	doi = {10.1088/0022-3719/10/7/008},
	language = {en},
	number = {7},
	urldate = {2025-08-14},
	journal = {Journal of Physics C: Solid State Physics},
	author = {Hirotsu, S.},
	year = {1977},
	pages = {967--985},
}

@article{cui_synthesis_2020,
	title = {Synthesis, {Crystal} {Structure} and {Photoelectric} {Response} of {All}‐{Inorganic} {Copper} {Halide} {Salts} {CsCuCl}$_{\textrm{3}}$},
	volume = {2020},
	issn = {1434-1948, 1099-0682},
	url = {https://chemistry-europe.onlinelibrary.wiley.com/doi/10.1002/ejic.202000168},
	doi = {10.1002/ejic.202000168},
	language = {en},
	number = {22},
	urldate = {2025-08-14},
	journal = {European Journal of Inorganic Chemistry},
	author = {Cui, S. and Chen, Y. and Tao, S. and Cui, J. and Yuan, C. and Yu, N. and Zhou, H. and Yin, J. and Zhang, X.},
	year = {2020},
	pages = {2165--2169},
}

@article{fernandez_new_1976,
	title = {New high temperature phase transitions in cesium cupric chloride},
	volume = {11},
	copyright = {https://www.elsevier.com/tdm/userlicense/1.0/},
	issn = {00255408},
	url = {https://linkinghub.elsevier.com/retrieve/pii/0025540876900167},
	doi = {10.1016/0025-5408(76)90016-7},
	language = {en},
	number = {9},
	urldate = {2025-08-14},
	journal = {Materials Research Bulletin},
	author = {Fernández, J. and Tello, M. J. and Peraza, J. and Bocanegra, E. H.},
	year = {1976},
	pages = {1161--1167},
}

@article{tanaka_structural_1986,
	title = {Structural phase transitions in hexagonal {ABCl}$_{\textrm{3}}$ {Jahn}-{Teller} crystals: {I}. {Face}-sharing coupling and ground-state configuration},
	volume = {19},
	issn = {0022-3719},
	shorttitle = {Structural phase transitions in hexagonal {ABCl}$_{\textrm{3}}$ {Jahn}-{Teller} crystals},
	url = {https://iopscience.iop.org/article/10.1088/0022-3719/19/25/008},
	doi = {10.1088/0022-3719/19/25/008},
	language = {en},
	number = {25},
	urldate = {2025-08-14},
	journal = {Journal of Physics C: Solid State Physics},
	author = {Tanaka, H. and Dachs, H. and Iio, K. and Nagata, K.},
	year = {1986},
	pages = {4861--4878},
}

@article{tanaka_structural_1986_part2,
	title = {Structural phase transitions in hexagonal {ABCl}$_{\textrm{3}}$ {Jahn}-{Teller} crystals: {II}. {Mean}-field approximation},
	volume = {19},
	issn = {0022-3719},
	shorttitle = {Structural phase transitions in hexagonal {ABCl3} {Jahn}-{Teller} crystals},
	url = {https://doi.org/10.1088/0022-3719/19/25/015},
	doi = {10.1088/0022-3719/19/25/015},
	language = {en},
	number = {25},
	urldate = {2026-07-06},
	journal = {Journal of Physics C: Solid State Physics},
	author = {Tanaka, H. and Dachs, H. and Iio, K. and Nagata, K.},
	year = {1986},
	pages = {4879},
}

@article{petzelt_farinfrared_1981,
	title = {Far‐infrared and {Raman} spectroscopy of the phase transition in {CsCuCl}$_{\textrm{3}}$},
	volume = {10},
	copyright = {http://onlinelibrary.wiley.com/termsAndConditions\#vor},
	issn = {0377-0486, 1097-4555},
	url = {https://analyticalsciencejournals.onlinelibrary.wiley.com/doi/10.1002/jrs.1250100137},
	doi = {10.1002/jrs.1250100137},
	language = {en},
	number = {1},
	urldate = {2025-08-14},
	journal = {Journal of Raman Spectroscopy},
	author = {Petzelt, J. and Gregora, I. and Vorlíček, V. and Fousek, J. and Březina, B. and Kozlov, G. V. and Volkov, A. A.},
	year = {1981},
	pages = {187--193},
}

@article{kousaka_crystal_2014,
	title = {Crystal {Growth} of {Chiral} {Magnetic} {Material} in {CsCuCl}$_{\textrm{3}}$},
	volume = {502},
	copyright = {http://iopscience.iop.org/info/page/text-and-data-mining},
	issn = {1742-6588, 1742-6596},
	url = {https://iopscience.iop.org/article/10.1088/1742-6596/502/1/012019},
	doi = {10.1088/1742-6596/502/1/012019},
	language = {en},
	urldate = {2025-08-14},
	journal = {Journal of Physics: Conference Series},
	author = {Kousaka, Y. and Koyama, T. and Miyagawa, M. and Tanaka, K. and Akimitsu, J. and Inoue, K.},
	year = {2014},
	pages = {012019},
}

@article{kroese_phase_1971,
    title = {A phase transition in a compound with helical electric dipole structure: {CsCuCl$_3$}},
    journal = {Solid State Communications},
    volume = {9},
    number = {19},
    pages = {1707-1709},
    year = {1971},
    issn = {0038-1098},
    doi = {https://doi.org/10.1016/0038-1098(71)90345-0},
    url = {https://www.sciencedirect.com/science/article/pii/0038109871903450},
    author = {C. J. Kroese and J. C. M. {Tindemans-van Eyndhoven} and W. J. A. Maaskant},
}

@article{kroese_high-temperature_1974,
	title = {The {High}-{Temperature} {Structure} of {CsCuCl$_3$}},
	language = {en},
	author = {Kroese, C. J. and Maaskant, W. J. A. and Verschoor, G. C.},
    journal = {Acta Crystallographica Section B: Structural Crystallography and Crystal Chemistry}, 
    year = {1974}, 
    volume = {30}, 
    pages = {1053-1056},    
}

@article{soboleva_growth_2008,
	title = {Growth of functional single crystals of complex compounds on the basis of the solubility phase diagrams of ternary systems},
	volume = {53},
	copyright = {http://www.springer.com/tdm},
	issn = {1063-7745, 1562-689X},
	url = {http://link.springer.com/10.1134/S1063774508010227},
	doi = {10.1134/S1063774508010227},
	language = {en},
	number = {1},
	urldate = {2025-08-14},
	journal = {Crystallography Reports},
	author = {Soboleva, L. V.},
	year = {2008},
	pages = {164--170},
}

@article{sorokin_electrical_2017,
	title = {Electrical conductivity of {CsCuCl$_3$} crystals at structural phase transition},
	volume = {62},
	issn = {1063-7745, 1562-689X},
	url = {http://link.springer.com/10.1134/S1063774517040253},
	doi = {10.1134/S1063774517040253},
	language = {en},
	number = {4},
	urldate = {2025-08-14},
	journal = {Crystallography Reports},
	author = {Sorokin, N. I.},
	year = {2017},
	pages = {629--631},
}

@article{van_well_mixed_2020,
	title = {Mixed system {Cs$_3$Cu$_3$Cl$_{8-x}$Br$_x$OH} with weakly connected {Cu}-triangles},
	volume = {140},
	issn = {00223697},
	url = {https://linkinghub.elsevier.com/retrieve/pii/S002236971932325X},
	doi = {10.1016/j.jpcs.2020.109386},
	language = {en},
	urldate = {2025-08-14},
	journal = {Journal of Physics and Chemistry of Solids},
	author = {Van Well, N. and Bolte, M. and Eisele, C. and Keller, L. and Schefer, J. and Van Smaalen, S.},
	year = {2020},
	pages = {109386},
}

@article{VASUDEVAN197944,
title = {Jahn-{Teller} effect induced phase transitions in {CsCuCl$_3$}},
journal = {Physics Letters A},
volume = {70},
number = {1},
pages = {44-46},
year = {1979},
issn = {0375-9601},
doi = {https://doi.org/10.1016/0375-9601(79)90322-0},
url = {https://www.sciencedirect.com/science/article/pii/0375960179903220},
author = {S. Vasudevan and A. M. Shaikh and C. N. R. Rao}
}

@article{forster_phonon_1997,
	series = {Proceedings of the {First} {European} {Conference} on {Neutron} {Scattering}},
	title = {Phonon renormalisation at small \textit{q} values in the {HT}-phase of {CsCuCl$_3$}},
	volume = {234-236},
	issn = {0921-4526},
	url = {https://www.sciencedirect.com/science/article/pii/S0921452696009295},
	doi = {10.1016/S0921-4526(96)00929-5},
	urldate = {2025-08-14},
	journal = {Physica B: Condensed Matter},
	author = {Förster, U. and Graf, H. A. and Schotte, U. and Stuhr, U.},
	year = {1997},
	pages = {142--143},
}

@article{oconnor_preparation_1970,
	title = {Preparation and properties of cesium cupric chloride},
	volume = {6},
	issn = {0022-0248},
	url = {https://www.sciencedirect.com/science/article/pii/0022024870900953},
	doi = {10.1016/0022-0248(70)90095-3},
	number = {4},
	urldate = {2025-08-14},
	journal = {Journal of Crystal Growth},
	author = {O'Connor, J. J. and Dipietro, M. A. and Armington, A. F.},
	year = {1970},
	pages = {346--348},
}

@article{tanaka_electron_1985,
	title = {Electron {Paramagnetic} {Resonance} in the {Quasi}-{One}-{Dimensional} {Jahn}-{Teller}-{Crystals}. {I}. {CsCuCl$_3$}},
	volume = {54},
	issn = {0031-9015},
	url = {https://journals.jps.jp/doi/abs/10.1143/JPSJ.54.4345},
	doi = {10.1143/JPSJ.54.4345},
	number = {11},
	urldate = {2025-08-14},
	journal = {Journal of the Physical Society of Japan},
	publisher = {The Physical Society of Japan},
	author = {Tanaka, H. and Iio, K. and Nagata, K.},
	year = {1985},
	pages = {4345--4358},
}

@article{graf_quasi-elastic_1989,
	title = {Quasi-elastic and inelastic neutron scattering in the high temperature phase of the cooperative {Jahn}-{Teller} system {CsCuCl}$_{\textrm{3}}$},
	volume = {1},
	issn = {0953-8984, 1361-648X},
	url = {https://iopscience.iop.org/article/10.1088/0953-8984/1/24/001},
	doi = {10.1088/0953-8984/1/24/001},
	language = {en},
	number = {24},
	urldate = {2025-08-18},
	journal = {Journal of Physics: Condensed Matter},
	author = {Graf, H. A. and Shirane, G. and Schotte, U. and Dachs, H. and Pyka, N. and Iizumi, M.},
	year = {1989},
	pages = {3743--3763},
}

@article{schotte_theory_1987,
	title = {Theory of diffuse neutron scattering in the high temperature phase of {CsCuCl$_3$} and related {Jahn}-{Teller} compounds},
	volume = {66},
	issn = {1431-584X},
	url = {https://doi.org/10.1007/BF01312766},
	doi = {10.1007/BF01312766},
	language = {en},
	number = {1},
	urldate = {2025-08-18},
	journal = {Zeitschrift für Physik B Condensed Matter},
	author = {Schotte, U.},
	year = {1987},
	pages = {91--101},
}

@article{graf_jahn-teller_1986,
	title = {Jahn-{Teller} phase transition in {CsCuCl$_3$} and {CsCrCl$_3$}: {A} neutron scattering study},
	volume = {57},
	issn = {0038-1098},
	shorttitle = {Jahn-{Teller} phase transition in {CsCuCl3} and {CsCrCl3}},
	url = {https://www.sciencedirect.com/science/article/pii/0038109886904953},
	doi = {10.1016/0038-1098(86)90495-3},
	number = {6},
	urldate = {2025-08-18},
	journal = {Solid State Communications},
	author = {Graf, H. A. and Tanaka, H. and Dachs, H. and Pyka, N. and Schotte, U. and Shirane, G.},
	year = {1986},
	pages = {469--472},
}

@article{kroese_relation_1974,
	title = {The relation between the high-temperature and room-temperature structure of {CsCuCl$_3$}},
	volume = {5},
	issn = {0301-0104},
	url = {https://www.sciencedirect.com/science/article/pii/0301010474800200},
	doi = {10.1016/0301-0104(74)80020-0},
	number = {2},
	urldate = {2025-08-18},
	journal = {Chemical Physics},
	author = {Kroese, C. J. and Maaskant, W. J. A.},
	year = {1974},
	pages = {224--233},
}

@article{soboleva_investigation_1976,
	title = {Investigation into some physical properties of the {CsCuCl$_3$} crystal},
	volume = {21},
	url = {https://inis.iaea.org/records/vwmb0-kfn57},
	language = {Russian},
	number = {6},
	urldate = {2025-10-16},
	author = {Soboleva, L. V. and Sil'vestrova, I. M. and Perekalina, Z. B. and Gil'varg, A. B. and Martyshev, Yu N.},
	year = {1976},
	pages = {1140--1147},
}

@article{qiu_orbital_2005,
	title = {Orbital {Correlations} in the {Pseudocubic} {$O$} and {Rhombohedral} {$R$} {Phases} of {LaMnO$_3$}},
	volume = {94},
	url = {https://link.aps.org/doi/10.1103/PhysRevLett.94.177203},
	doi = {10.1103/PhysRevLett.94.177203},
	number = {17},
	urldate = {2025-12-10},
	journal = {Physical Review Letters},
	publisher = {American Physical Society},
	author = {Qiu, X. and Proffen, Th. and Mitchell, J. F. and Billinge, S. J. L.},
	year = {2005},
	pages = {177203},
}

@incollection{maaskant_helices_1995,
	address = {Berlin, Heidelberg},
	title = {On helices resulting from a cooperative {Jahn}-{Teller} effect in hexagonal perovskites},
	isbn = {978-3-540-49188-0},
	url = {https://doi.org/10.1007/3-540-59105-2_2},
	doi = {10.1007/3-540-59105-2_2},
	language = {en},
	urldate = {2026-02-05},
	booktitle = {Iron-{Sulfur} {Proteins} {Perovskites}},
	publisher = {Springer},
	author = {Maaskant, W. J. A.},
	year = {1995},
	pages = {55--87},
}

@article{haije_magnetic_1986,
	title = {A magnetic susceptibility study on the dynamic {Jahn}-{Teller} effects in {CsCuCl$_3$}: single octahedron and cooperative phenomena},
	volume = {19},
	issn = {0022-3719},
	shorttitle = {A magnetic susceptibility study on the dynamic {Jahn}-{Teller} effects in {CsCuCl3}},
	url = {https://doi.org/10.1088/0022-3719/19/35/008},
	doi = {10.1088/0022-3719/19/35/008},
	language = {en},
	number = {35},
	urldate = {2026-02-13},
	journal = {Journal of Physics C: Solid State Physics},
	author = {Haije, W. G. and Maaskant, W. J. A.},
	year = {1986},
	pages = {6943},
}

@article{schotte_elastic_1989,
	title = {Elastic dipole theory of {CsCuCl$_3$} and related hexagonal {Jahn}-{Teller} compounds for neutron scattering},
	volume = {1},
	issn = {0953-8984},
	url = {https://doi.org/10.1088/0953-8984/1/24/002},
	doi = {10.1088/0953-8984/1/24/002},
	language = {en},
	number = {24},
	urldate = {2026-02-13},
	journal = {Journal of Physics: Condensed Matter},
	author = {Schotte, U. and Graf, H. A. and Dachs, H.},
	year = {1989},
	pages = {3765},
}

@article{crama_jahnteller_1981,
	title = {The {Jahn}–{Teller} distorted structure of caesium copper({II}) trichloride},
	volume = {37},
	issn = {0567-7408},
	url = {https://journals.iucr.org/b/issues/1981/12/00/a20407/},
	doi = {10.1107/S0567740881008224},
	language = {en},
	number = {12},
	urldate = {2026-03-26},
	journal = {Acta Crystallographica Section B: Structural Crystallography and Crystal Chemistry},
	publisher = {International Union of Crystallography},
	author = {Crama, W. J.},
	year = {1981},
	pages = {2133--2136},
}

@article{schlueter_redetermination_1966,
	title = {A {Redetermination} of the {Crystal} {Structure} of {CsCuCl$_3$}},
	volume = {5},
	issn = {0020-1669},
	url = {https://doi.org/10.1021/ic50036a025},
	doi = {10.1021/ic50036a025},
	number = {2},
	urldate = {2026-05-01},
	journal = {Inorganic Chemistry},
	publisher = {American Chemical Society},
	author = {Schlueter, A. W. and Jacobson, R. A. and Rundle, R. E.},
	year = {1966},
	pages = {277--280},
}

@article{nagle-cocco_displacive_2024,
	title = {Displacive {Jahn}–{Teller} {Transition} in {NaNiO$_2$}},
	volume = {146},
	issn = {0002-7863},
	url = {https://doi.org/10.1021/jacs.4c09922},
	doi = {10.1021/jacs.4c09922},
	number = {43},
	urldate = {2026-02-13},
	journal = {Journal of the American Chemical Society},
	publisher = {American Chemical Society},
	author = {Nagle-Cocco, L. A. V. and Genreith-Schriever, A. R. and Steele, J. M. A. and Tacconis, C. and Bocarsly, J. D. and Mathon, O. and Neuefeind, J. C. and Liu, J. and O’Keefe, C. A. and Goodwin, A. L. and Grey, C. P. and Evans, J. S. O. and Dutton, S. E.},
	year = {2024},
	pages = {29560--29574},
}

@article{dey_monoclinic_2022,
	title = {Monoclinic symmetry at the nanoscale in lead-free ferroelectric {BaZr$_x$Ti$_{1-x}$O$_3$} ceramics},
	volume = {105},
	url = {https://link.aps.org/doi/10.1103/PhysRevB.105.174202},
	doi = {10.1103/PhysRevB.105.174202},
	number = {17},
	urldate = {2026-02-16},
	journal = {Physical Review B},
	publisher = {American Physical Society},
	author = {Dey, K. and Tripathy, A. and Sahu, S. R. and Srivastava, H. and Sagdeo, A. and Strempfer, J. and Shukla, D. K.},
	year = {2022},
	pages = {174202},
}

@article{shoemaker_unraveling_2009,
	title = {Unraveling {Atomic} {Positions} in an {Oxide} {Spinel} with {Two} {Jahn}-{Teller} {Ions}: {Local} {Structure} {Investigation} of {CuMn$_2$O$_4$}},
	volume = {131},
	issn = {0002-7863},
	shorttitle = {Unraveling {Atomic} {Positions} in an {Oxide} {Spinel} with {Two} {Jahn}−{Teller} {Ions}},
	url = {https://doi.org/10.1021/ja902096h},
	doi = {10.1021/ja902096h},
	number = {32},
	urldate = {2026-06-23},
	journal = {Journal of the American Chemical Society},
	publisher = {American Chemical Society},
	author = {Shoemaker, D. P. and Li, J. and Seshadri, R.},
	year = {2009},
	pages = {11450--11457},
}

@article{jiang_probing_2021,
	title = {Probing the {Local} {Site} {Disorder} and {Distortion} in {Pyrochlore} {High}-{Entropy} {Oxides}},
	volume = {143},
	issn = {0002-7863},
	url = {https://doi.org/10.1021/jacs.0c10739},
	doi = {10.1021/jacs.0c10739},
	number = {11},
	urldate = {2026-06-23},
	journal = {Journal of the American Chemical Society},
	publisher = {American Chemical Society},
	author = {Jiang, B. and Bridges, C. A. and Unocic, R. R. and Pitike, K. C. and Cooper, V. R. and Zhang, Y. and Lin, D. and Page, K.},
	year = {2021},
	pages = {4193--4204},
}

\end{document}